\documentclass[aip,jcp,amsmath,amssymb,longbibliography,floatfix,reprint]{revtex4-2}
\usepackage{graphicx}
\usepackage{dcolumn}
\usepackage{enumitem}
\usepackage{float}
\usepackage{bm}
\usepackage[export]{adjustbox}
\usepackage{hyperref}
\usepackage[mathlines]{lineno}
\usepackage[utf8]{inputenc}
\usepackage[T1]{fontenc}
\usepackage{etoolbox}
\usepackage{siunitx}
\usepackage{chemformula}
\usepackage{tikz}
\usetikzlibrary{quantikz2}

\newcommand{\vectheta}{{\bm{\theta}}}
\makeatletter
\def\@email#1#2{%
 \endgroup
 \patchcmd{\titleblock@produce}
  {\frontmatter@RRAPformat}
  {\frontmatter@RRAPformat{\produce@RRAP{*#1\href{mailto:#2}{#2}}}\frontmatter@RRAPformat}
  {}{}
}%
\makeatother
\begin{document}
\title{Qubit encodings for lattices of dipolar planar rotors}
\author{Muhammad Shaeer Moeed}
\affiliation{ 
Department of Chemistry, University of Waterloo, Waterloo, Ontario N2L 3G1, Canada
}
\affiliation{ 
Institute For Quantum Computing, University of Waterloo, Waterloo, Ontario N2L 3G1, Canada
}
\affiliation{ Perimeter Institute for Theoretical Physics, Waterloo, Ontario N2L 2Y5, Canada}
\author{James Brown}
\affiliation{ 
qBraid Co., 111 S Wacker Drive, Chicago, IL, USA
}
\author{Alexander Ibrahim}
\affiliation{ 
Department of Physics, University of Waterloo, Waterloo, Ontario N2L 3G1, Canada
}
\author{Estevao Vilas Boas De Oliveira}
\affiliation{ 
Department of Physics, University of Waterloo, Waterloo, Ontario N2L 3G1, Canada
}
\affiliation{ 
Institute For Quantum Computing, University of Waterloo, Waterloo, Ontario N2L 3G1, Canada
}
\author{Pierre-Nicholas Roy}
\affiliation{ 
Department of Chemistry, University of Waterloo, Waterloo, Ontario N2L 3G1, Canada}
\affiliation{ 
Institute For Quantum Computing, University of Waterloo, Waterloo, Ontario N2L 3G1, Canada
}
\affiliation{ Perimeter Institute for Theoretical Physics, Waterloo, Ontario N2L 2Y5, Canada}
\email{pnroy@uwaterloo.ca}

\date{\today}

\begin{abstract}
    Near term quantum devices have recently garnered significant interest as promising candidates for investigating difficult-to-probe regimes in many-body physics. To this end, various qubit encoding schemes targeting second quantized Hamiltonians have been proposed and optimized. In this work, we investigate two qubit representations of the planar rotor lattice Hamiltonian. The first representation is realized by decomposing the rotor Hamiltonian projectors in binary and mapping them to spin-$1/2$ projectors. The second approach relies on embedding the planar rotor lattice Hilbert space in a larger space and recovering the relevant qubit encoded system as a quotient space projecting down to the physical degrees of freedom. This is typically called the unary mapping and is used for bosonic systems. We establish the veracity of the two encoding approaches using sparse diagonalization on small chains and discuss quantum phase estimation resource requirements to simulate small planar rotor lattices on near-term quantum devices. 
\end{abstract}

\maketitle

\section{Introduction}

Rotor lattices offer a rich and versatile theoretical landscape for investigating the emergence of many-body phases in confined molecular assemblies.\cite{endohedral_fullerites_exp, exp_nanotube} A pertinent example is the confinement of water molecules in crystals such as Beryl and Cordierite.\cite{belyanchikov_2020_dielectric} Experimental results have indicated that the confining crystal structure screens the usual hydrogen bonding between water molecules and allows them to interact via their dipolar coupling when confined to such lattices. \cite{gorshunov_2016_incipient} This allows for the existence of an ordered phase at low temperatures causing the assembly to exhibit ferro-electricity. It has recently been suggested that this behavior can be modeled using a $3$ dimensional lattice of planar rotors with a dipolar coupling between different sites.\cite{gorshunov_2016_incipient} 

Moreover, $2$ dimensional lattices of planar rotors have also been used to model the orientational ordering of homo-nuclear diatomic molecules adsorbed on inert surfaces.\cite{Berne, freiman20102d} This so-called Anisotropic Planar Rotor (APR) model and its quantum variant,\cite{Berne} assume that the diatomic molecules perform uni-axial rotation with fixed bond lengths while interacting with a quadrupolar 
interaction. Computational studies targeting this model have suggested the possibility of a re-entrant disordered phase at low temperatures in these systems.\cite{freiman20102d, hetenyi2005orientational} This is interesting from a theoretical perspective because it underscores the need for the development of robust algorithmic tools capable of resolving the phase diagram in higher dimensional lattices as well as lower ones. 

Discretized planar rotors are typically studied in the context of the quantum clock model.\cite{baxter1989simple, baxter1989superintegrable} It has been suggested that a $1$ dimensional lattice of chiral clocks interacting with a dot-product coupling may admit topological order in the form of parafermionic edge zero modes.\cite{fendley2012parafermionic} Experimentally, critical points of such clock models can be realized Rydberg atom arrays .\cite{bernien2017probing, whitsitt2018quantum} The phase diagram of $1$ dimensional quantum clock lattices have been investigated in the past using theoretical tools as well as Density Matrix Renormalization Group (DMRG)\cite{DMRG_Review_Article, dmrg} studies, as the order of discretization is varied.\cite{whitsitt2018quantum} Finally, $2$ dimensional lattices of planar rotors with a $U(1)$ symmetry (the classical XY model) have been famously explored in the context of Berezinskii–Kosterlitz–Thouless (BKT) phase transitions in past work.\cite{kosterlitz1973ordering} 

Since Exact Diagonalization (ED) scales exponentially with system sizes, various polynomially scaling computational techniques have been developed to study the many-body statistical mechanics of planar rotor lattices. These include DMRG\cite{planar_rotor_QPT} as well as stochastic methods such as Path Integral Monte Carlo (PIMC).\cite{moeed2025pair, zhang2025path, patil2021unconventional, marx1999path} However, current DMRG capabilities targeting planar rotations are limited to $1$ dimensional lattices. Stochastic methods generally suffer from critical slowing down near phase transitions\cite{wolff1989critical, moeed2025pair} as well as the sign problem which limits their utility in the context of frustrated systems.\cite{iglovikov2015geometry} 

In recent years, quantum devices have been proposed as promising candidates for simulating the many-body physics of spin systems, as well as those involving fermionic and bosonic excitations.\cite{Somma_QC, kim2010quantum, alexeev2025perspective} To this end, various qubit encoding schemes have been developed to study electronic structure problems in quantum chemistry such as the Jordan-Wigner mapping and the Bravyi-Kitaev transformation.\cite{seeley2012bravyi, li2022unified} However, these approaches can not be readily used to simulate systems that are naturally described by first quantized Hamiltonians such as planar rotor lattices. In this work, we develop two different qubit encoding for mapping an $L$-dimensional lattice of planar rotors to a $(L+1)$-dimensional lattice of spin-1/2 particles. We computationally verify both approaches via ED studies targeting small nearest-neighbor chains. Then, we perform Quantum Phase Estimation (QPE)\cite{nielsenchuang, mande2023tightboundsquantumphase} and Variational Quantum Eigensolver (VQE)\cite{HVAVQE, TFIMVQE} simulations to explore the resource requirements for simulating such systems on quantum devices. 

This paper is organized as follows: in Sec. \ref{theory}, we develop the qubit mapped Hamiltonians corresponding to a lattice of dipolar planar rotors in both encoding schemes. Following that, in Sec. \ref{algorithms}, we discuss our QPE framework as well as our VQE ansatz. Then, in Sec. \ref{section_3}, we simulate the performance of quantum algorithms classically and discuss our results. Sec. \ref{conclusions_outlook} is dedicated to conclusions and future directions. 

\section{Theoretical Framework} \label{theory}

\subsection{Planar Rotor Lattice}

In this work, we consider an $L$-dimensional lattice ($L \in \{1,2,3\})$ of planar rotors with a dipole-dipole interaction. The system is defined by the Hamiltonian,\cite{zhang2025path}
\begin{equation}
    H = B\sum_{i=1}^{N} l_i^2 + \sum_{i<j}^{N_b} V_{i,j}
\end{equation}
where $l_i$ is the angular momentum of a single rotor and $V_{ij}$ is the dipole interaction between sites $i$ and $j$. $N_b$ here denotes the number of bonds in the rotor lattice. In a general system, this is the number of pairs of rotors: $N_b = \binom{N}{2}$. The pair-wise interaction is given by,\cite{jackson_classical_1999}
\begin{equation}
    V_{ij} = \frac{g_0}{R_{ij}^3} (e_i \cdot e_j - 3 (e_i \cdot r_{ij}) (e_j \cdot r_{ij}))
\end{equation}
Here, $g_0 = \mu_0/4\pi \epsilon_0$ is a site-independent constant, $R_{ij}$ is the distance between sites $i$ and $j$, $e_i$ is the dipole moment corresponding to site $i$ and $r_{ij}$ is the unit vector connecting sites $i$ and $j$. Expressing the dipole moment for each site in components ($e_i = (x_i, y_i)$), we get the following interaction for a general 3 dimensional lattice, 
\begin{gather}
    V_{ij} = \frac{g_0}{R_{ij}^3} [(1-3\cos^2(\gamma_{ij}^1)\sin^2(\gamma_{ij}^{2}))x_ix_j \nonumber \\ + (1-3\sin^2(\gamma_{ij}^1)\sin^2(\gamma_{ij}^{2})) y_iy_j \nonumber \\ - 3 \cos(\gamma_{ij}^{1})\sin(\gamma_{ij}^{1})\sin^2(\gamma_{ij}^{2}) (x_i y_j + x_j y_i)]
\end{gather}
Here, $\gamma_{ij}^{1}$ is the polar angle and the $\gamma_{ij}^{2}$ is the azimuthal angle associated with the vector $r_{ij}$. For a coplanar chain of rotors with a pairwise nearest neighbor interaction, we get,\cite{planar_rotor_QPT, moeed2025pair}
\begin{equation}
    V = \sum_{i=1}^{N-1}(y_iy_j - 2x_ix_j)
\end{equation}
Since each dipole moment is a unit vector, we can parameterize it by a polar angular coordinate $\phi$ to get $x_i = \cos(\phi_i)$ and $y_i = \sin(\phi_i)$. In the momentum representation, the diagonal angular momentum operator admits a discrete basis due to periodic boundaries in position space,\cite{estevao_pimc}
\begin{gather}
    l_i = \sum_{m_i=-\infty}^{\infty} m_i |m_i\rangle \langle m_i| = \sum_{m_i=-l}^{n} m_i |m_i\rangle \langle m_i|
\end{gather}
\begin{figure}[b]
    \centering
    \includegraphics[width=\linewidth]{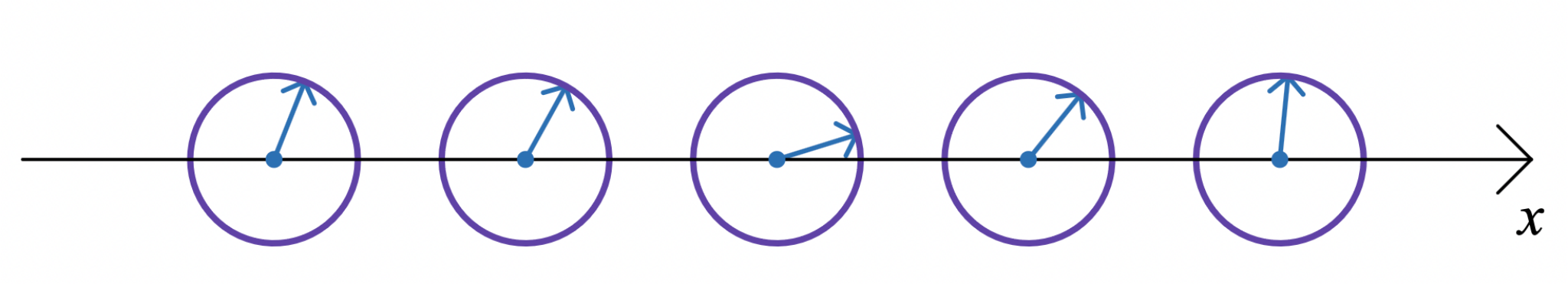}
    \caption{Schematic representation of a coplanar chain of planar rotors}
    \label{coplanar_chain_rep}
\end{figure}
Because we are interested in studying the ground state properties of such systems computationally, we have introduced a basis set cut-off, $-l \leq m \leq n$, in the momentum representation as the higher energy single rotor states do not contribute appreciably to the many-body ground state.\cite{Meyer_DVR} In particular, for $1$d chains, it has been shown that using $m \in [-4, 4]$ is sufficient for convergence via DMRG studies in previous work.\cite{planar_rotor_QPT} To construct the interaction term we use the momentum representation of the position operators which can be expressed as linear combinations of the increment and decrement operators ($S^+_{j}$ and $S_j^-$),\cite{nielsenchuang, increment_decrement}
\begin{gather}
    x_j = \frac{1}{2}(S_j^{+} + S_j^{-}) = \frac{1}{2} \sum_{m_j=-l}^{n}|m_j+1\rangle \langle m_j| + h.c. \label{x_j_momentum} \\ y_j =  \frac{1}{2i}(S_j^{+} - S_j^{-}) = \frac{1}{2i} \sum_{m_j=-l}^{n} |m_j+1\rangle \langle m_j| + h.c. \label{y_j_momentum}
\end{gather}
Here, the state $|n+1\rangle$ is identified with $|-l\rangle$ and $|-l-1\rangle$ is identified with $|n\rangle$ so that the momentum states obey periodic boundary conditions. This is equivalent to constructing a Fourier series for band-limited functions by taking copies of the frequency spectrum and has the effect of imposing a discretization in position space. The interaction operators $x_ix_j$, $y_iy_j$ and $(x_iy_j + x_jy_i)$ are then given by, 
\begin{gather}
    x_ix_j = \frac{1}{4}\sum_{m_{i}, m_{j}}|m_{i} + 1, m_{j} - 1\rangle \langle m_i, m_j | \nonumber \\ + |m_i + 1, m_j + 1\rangle \langle m_i, m_j| + h.c. \label{xixj_1} \\
    y_iy_j = \frac{1}{4}\sum_{m_{i}, m_{j}}|m_{i} + 1, m_{j} - 1\rangle \langle m_i, m_j | \nonumber \\ - |m_i + 1, m_j + 1\rangle \langle m_i, m_j| + h.c. \label{yiyj_1} \\
    (x_iy_j + x_jy_i) = \frac{i}{2} \sum_{m_i, m_j} |m_{i}-1, m_{j}-1\rangle \langle m_i, m_j| \nonumber \\ + h.c. \label{cross_1}
\end{gather}
For our computational analysis, we use a  coplanar chain with nearest-neighbor interactions as shown in Fig. \ref{coplanar_chain_rep}. The Hamiltonian (with a momentum cutoff) for that system is given by,   
\begin{gather}
    H = \sum_{i=1}^{N} K_i + g\sum_{i=1}^{N-1} V_{i,i+1} \\ K_i = \sum_{m_{i}}^{d-1} (m_{i} - l)^2 |m_i\rangle \langle m_i| \label{kinetic_energy_momentum_rep} \\ V_{i,i+1} = - \frac{1}{4}\sum_{m_i,m_{i+1}}^{d-1} \bigg(|m_{i} + 1, m_{j} - 1\rangle \langle m_i, m_j|\bigg) \nonumber \\ + \frac{3}{4}\sum_{m_i,m_{i+1}}^{d-1} \bigg(|m_{i} + 1, m_{j} - 1\rangle \langle m_i, m_j|\bigg) + h.c. \label{rotor_momentum_representation_coplanar}
\end{gather}
Note that the terms $m_i+1$ and $m_i-1$ here correspond to addition and subtraction modulo $d = n+l+1$, the number of states because we use periodic boundaries in momentum space. Moreover, we have transformed the sum in $K_i$ such that $m_i \rightarrow m_i + l$ for convenience. 

This system has been studied in recent work via DMRG\cite{planar_rotor_QPT} as well as Path Integral Ground State (PIGS) simulations.\cite{zhang2025path, moeed2025pair} The Hamiltonian in Eq. \eqref{rotor_momentum_representation_coplanar} has a $Z_2$ symmetry associated with rotating all rotors in the chain by an angle $\pi$. In the thermodynamic limit, it has been shown that this symmetry is spontaneously broken at $g \approx 0.5$ resulting in a $(1+1)$-dimensional quantum phase transition that is in the same universality class as the Transverse Field Ising Model (TFIM).\cite{planar_rotor_QPT} For $g < 0.5$, the system then admits a disordered phase with the ground state being given by $|E_0\rangle = \bigotimes_i |m_i=0\rangle$. As $g$ to increased to $0.5$, the system enters an ordered phase with all the rotors aligned along the chain. Here the exact ground state is given by: 
\begin{equation}
    |E_0\rangle  = \frac{1}{\sqrt{2}}\left(\bigotimes_i|\phi_i = 0\rangle + \bigotimes_i|\phi_i = \pi \rangle\right) \label{ground_state_large_g}
\end{equation}
It should be noted that when $g$ is small, even for small chains, the ground state is a product state in the momentum representation and can be easily mapped to a product state in the qubit representations we develop. However, as $g \rightarrow \infty$, the ground state approaches Eq. \eqref{ground_state_large_g} and no longer has this property.\cite{estevao_pimc} Consequently, even though, we restrict ourselves in this work to small chains, we will computationally simulate both of these regimes independently to analyze the behavior of the approaches we develop here. 

\subsection{Binary Qubit Encoding} \label{binary_encoding}
Here, we introduce the binary encoding\cite{sawaya2020resource, mcardle2019digital} for the dipolar planar rotor chain. Since we have a finite basis set (in the momentum representation), we can construct a qubit mapping by representing each single rotor state $|m\rangle$ for $-l \leq m \leq n$ in binary. For $d=n+l+1$ states, we need $k=\lceil \log_2(d) \rceil$ qubits to construct a unique binary representation of each single-rotor state. This yields for the single rotor kinetic energy operator,
\begin{gather}
    l_{j}^2 = \sum_{m_j = -l}^{n} m_j^2 |m_j\rangle\langle m_j| =  \sum_{m_j = 0}^{d-1} (m_j - l)^2 |m_j\rangle \langle m_j| \nonumber \\ = \sum_{i_1^j, ..., i_k^j} \left(\sum_{r=1}^{k} 2^{k-r} i_{r}-l\right)^2 |i_1^j,...,i_k^j\rangle \langle i_1^j, ..., i_k^j| \label{qbit_rep_1_diagonal}
\end{gather}
where we have re-expressed the eigenvalues $(m_j-l)^2$ using the binary representation of $m_j \in \{0, ..., d-1\}$, 
\begin{equation}
    m_j = \sum_{r=1}^{k} 2^{k-r} i_{r}^j
\end{equation}
To construct the qubit operators, we can expand the square of the eigenvalues of $l_j$ in Eq. \eqref{qbit_rep_1_diagonal}. Then, the fact that $i_r \in \{0,1\}$ implies that the terms corresponding to $i_r = 0$ vanish. Expressing the projectors as $|1\rangle \langle 1| = \frac{1}{2}(I - \sigma_z)$ and restricting $d$ to be a power to $2$, we get (see Supplementary Materials Sec. A for details),
\begin{gather}
    l_{j}^{2} = \left(\frac{n+l+1}{2}\right)^2 \sum_{r,s} 2^{-r-s} \sigma_{r,j}^{z} \sigma_{s,j}^{z} \nonumber \\ - \frac{(n+l+1)(n-l)}{2} \sum_{r=1}^{k} 2^{-r} \sigma^{z}_{r,j} + \frac{1}{4}(n-l)^2 \label{single_rotor_ke_first_map}
\end{gather}
It is instructive to note here that while Eq. \eqref{single_rotor_ke_first_map} corresponds to the kinetic energy of a single rotor, it contains $2$-body Ising-like interaction terms between the spins used to represent the rotor basis states. These result from the squaring of the angular momentum operator as can be seen by factoring Eq. \eqref{single_rotor_ke_first_map},
\begin{equation}
    l_{j}^2 = \left(\frac{n+l+1}{2} \left(\sum_{r=1}^{k} 2^{-r} \sigma_{r,j}^{z} - \frac{n-l}{n+l+1}\right)\right)^2
\end{equation}
There is also a longitudinal field in the kinetic energy operator (using the Ising model parlance). This ensures that the ground state is a product state corresponding to all qubits in state $|0\rangle$ as would be expected from the definition in Eq. \eqref{kinetic_energy_momentum_rep}. For the potential energy operators, it is sufficient to construct the binary representation of the single rotor increment and decrement terms ($S_{j}^{+}$ and $S_{j}^{-}$) since the many-body potential is always a linear combination of products of those operators. 
Representing $m_j$ using $k$ bits, we get $|m_j\rangle = |i_1^j,...,i_k^j\rangle$. We can then construct the binary representation of $|m_j + 1\rangle$, by flipping all the qubits in the binary representation of $|m_j\rangle$ up to and including the first $|0\rangle$ starting from the right. Therefore, the increment operator is given by, 
\begin{gather}
    S^+_j = \sum_{m_j=0}^{d-1}|m_j+1\rangle \langle m_j| \nonumber \\ =\sum_{r=1}^{k} \sigma_{r,j}^{+} \left(\prod_{s=r+1}^{k}\sigma_{s,j}^{-}\right) + \left(\prod_{s=1}^{k}\sigma_{s,j}^{-}\right) \label{binary_mapping_S_plus}
\end{gather}
Here, the operators $\sigma_{r,j}^{\pm}$ are defined as $\frac{1}{2}(\sigma_{r,j}^{x} \pm  i \sigma_{r,j}^{y})$. It is instructive to note that for an arbitrary qubit string basis state $|i_1^j, ..., i_k^j\rangle$ with $i_a = 0$ and $i_b = 1$ for all $b > a$, the action of the above sum of Pauli strings is $0$ for all terms except for the operator corresponding to $r=a$ because $\sigma_{r,j}^{+}|1\rangle = \sigma_{r,j}^{-}|0\rangle = 0$. The final term in the sum in Eq. \eqref{binary_mapping_S_plus}, which contains no $\sigma^{+}_{r,j}$ operators is equivalent to $|-l\rangle \langle n| = |0^k\rangle \langle 1^k|$ arising due to the modulo $d$ addition and subtraction in Eqs. \eqref{x_j_momentum} and \eqref{y_j_momentum}. The decrement operator can similarly be constructed by replacing $\sigma^-$ with $\sigma^+$ in Eq. \eqref{binary_mapping_S_plus}. Finally, the operator $x_j$ (given by Eq. \eqref{x_j_momentum}), can then be expressed in this representation as follows,
\begin{gather}
    x_j = \frac{1}{2}\sum_{r=1}^{k} \left(\sigma_{r,j}^{+} \prod_{s=r+1}^{k}\sigma_{s,j}^{-} + \sigma_{r,j}^{-} \prod_{s=r+1}^{k}\sigma_{s,j}^{+}\right) \nonumber \\ + \frac{1}{2}\prod_{s=1}^{k}\sigma_{s,j}^{-} + \frac{1}{2}\prod_{s=1}^{k}\sigma_{s,j}^{+} 
\end{gather}
We can also express this as a sum over Pauli strings using the definitions of $\sigma^{\pm}$ (see Supplementary Materials Sec. B for details), 
\begin{gather}
    x_{j} = \frac{1}{2}\sum_{r=1}^{k} \frac{1}{2^{k-r}} \Biggl( \sum_{\gamma \ \text{even}} (-1)^{n_y/2} \sigma_{r,j}^{x} \prod_{s=r+1}^{k} \sigma_{r,j}^{\gamma(s)}  \nonumber \\ + \sum_{\gamma \ \text{odd}} (-1)^{(n_y-1)/2} \sigma_{r,j}^{y}\prod_{s=r+1}^{k} \sigma_{s,j}^{\gamma(s)}\Biggr) \nonumber \\ + \frac{1}{2^k} \sum_{\gamma \ \text{even}} (-1)^{n_y/2} \prod_{s=1}^{k} \sigma_{s,k}^{\gamma(s)}
\end{gather}
Here, $\gamma$ is an index string such that $\gamma = (\gamma(r+1), ..., \gamma(k))$ with $\gamma(s) \in \{x,y\}$ and  $n_y$ is the number of indices $s \in \{r+1, s\}$ such that $\gamma(s) = y$. In the above equation $\gamma \ \text{even}$ denotes index strings with an even number of elements equaling $y$ and vice versa for $\gamma \ \text{odd}$. The last term comes from operator strings corresponding to products of only $\sigma^{+}$ and products of only $\sigma^{-}$ terms in the definition of the increment and decrement operators. The operator $y_j$ can be constructed similarly.

We can now construct the required potential operators as specified in Eqs. \eqref{xixj_1}, \eqref{yiyj_1} and \eqref{cross_1} to simulate any lattice of planar rotors.  For a coplanar chain the potential term is a linear combination of $x_ix_j$ and $y_iy_j$. Since the potential is quadratic in position, and because the largest Pauli weight for a single rotor position operator is $k$, the largest Pauli string we will have for a $k$ bit representation is $2k$ for any planar rotor lattice in this encoding. 

\subsection{Bosonic Qubit Encoding} \label{bosonic_encoding}

The binary encoding discussed in section $\ref{binary_encoding}$ maps the on-site kinetic energy operator to a quadratic Hamiltonian. Therefore, the single rotor kinetic energy operator in that representation has spin-spin interaction terms. Moreover, the weight of the Pauli strings in the dipole-dipole interaction increases linearly with $k=\log_2(n+l+1)$. In this section, we construct the bosonic encoding (also referred to as the one-hot or unary mapping)\cite{Somma_QC, sawaya2019quantum, DiMatteo_one_hot} that yields a constant interaction Pauli weight and results in a kinetic energy term linear in spin operators. However, the cost associated with this simplicity is a larger qubit Hilbert space; this encoding requires $d=n+l+1$ qubits to represent each rotor. These extra degrees of freedom later need to be projected out to recover the many-body rotor Hilbert space from the embedding. To construct this mapping, first note that the angular momentum operator for this Hamiltonian can be identified with the shifted bosonic number operator (truncated appropriately), 
\begin{gather}
    l_j = \sum_{m_j}^{d} (m_j - l) |m_j\rangle \langle m_j| = (n_j - l) \\
    n_j |k\rangle = k |k\rangle
\end{gather}
Therefore, we can use the typical qubit encoding for bosonic modes\cite{Somma_QC} to map this system to a lattice of spin-1/2 particles as well. To this end, we identify the single rotor momentum basis states $|m\rangle$ for $1\leq m \leq d$ with a $d$-qubit product state, 
\begin{gather}
    |m\rangle \rightarrow |i_0,...,i_{d-1}\rangle, \ \ \ i_k = \delta_{m,k}
\end{gather}
\begin{figure}
    \centering
    \includegraphics[scale=0.3]{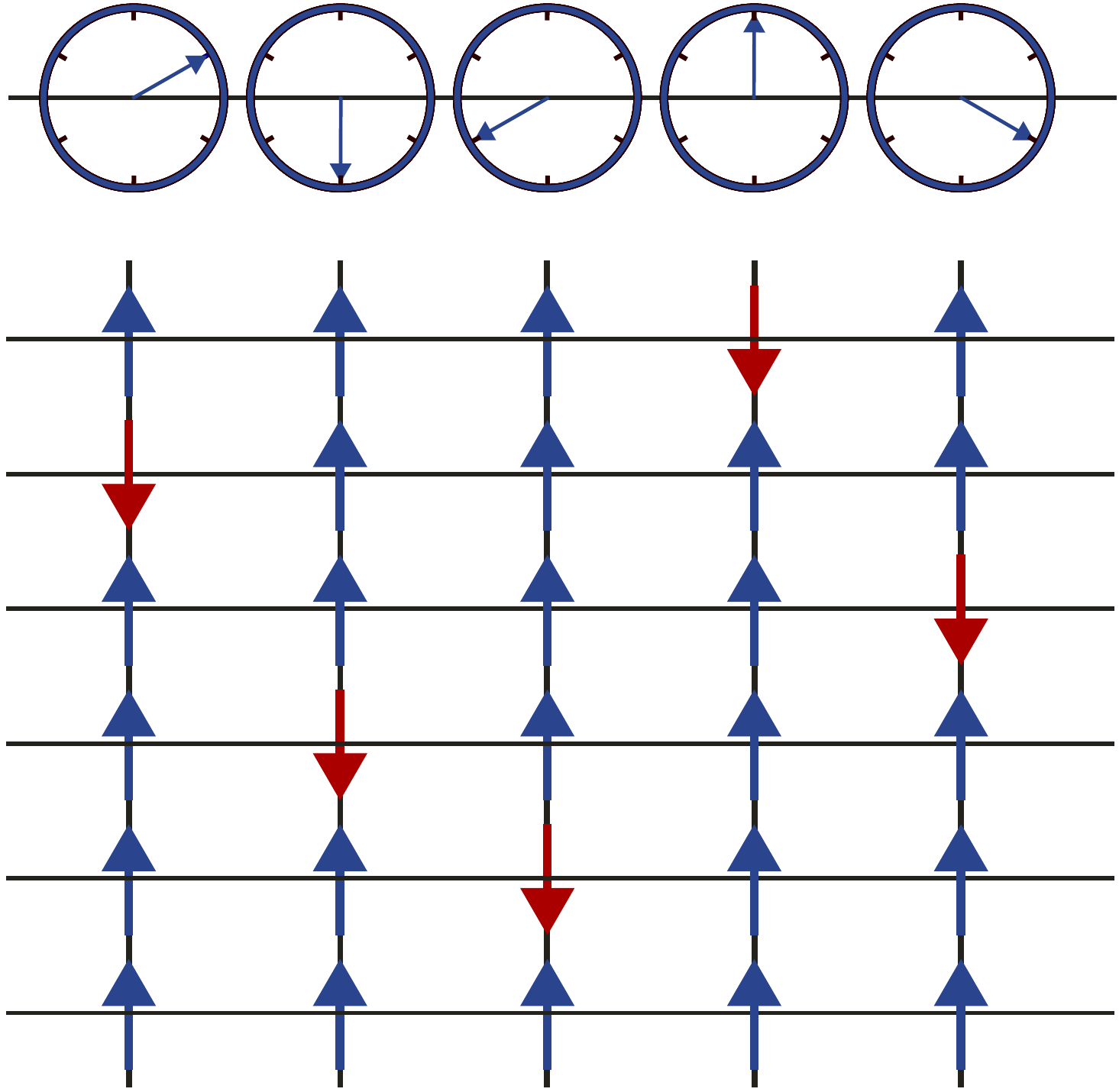}
    \caption{Schematic representation of a chain of planar rotors encoded into a lattice of spins using the bosonic mapping approach. Each of the ticks on the clocks represent different momentum eigenstates. The spin-down states correspond to the state $|1\rangle$ and the spin-up states correspond to the state $|0\rangle$ in the qubit encoding. A single rotor momentum eigenstate is mapped to a constrained chain of spins with exactly one excited spin site.}
    \label{sparse_map}
\end{figure}
where the $m$th qubit is in state $|1\rangle$ and the remaining $d-1$ qubits are in state $|0\rangle$. This state encoding is exhibited in Fig. \ref{sparse_map}. Using this identification, the projection operators $|m_j\rangle\langle m_j|$ needed for the truncated bosonic number operator become,
\begin{gather}
    |m_j\rangle\langle m_j| = |0,...,1,...,0\rangle \langle 0,...,1,...,0| \nonumber \\ = \frac{1}{2^d} (I + \sigma_{0,j}^{z})...(I - \sigma_{m_j,j}^{z})...(I+\sigma_{d-1,j}^z) 
\end{gather}
The kinetic energy operator in this representation can then be expressed as, 
\begin{equation}
    l_j^2 = \sum_{m_j}^{d-1} \frac{(m_j - l)^2}{2^d} (I - \sigma_{m_j,j}^{z}) \prod_{k \neq m_j}^{d-1}(I + \sigma_{k,j}^{z}) \label{bosonic_mapping_ke_1}
\end{equation}
The qubit Hilbert space we have mapped to here is $\mathcal{H}_Q = (\mathbb{C}^{2})^{\otimes d}$ with dimension $2^{d}$ while our single rotor system Hilbert space $\mathcal{H}_S$ has dimension $d$. This discrepancy arises because only a small number of vectors in $\mathcal{H}_Q$ are relevant physical states for our rotor system. The extra degrees of freedom correspond to the set $\mathcal{H}_V \subset \mathcal{H}_Q$ defined by the span of all product states with more than one qubit in state $|1\rangle$, and the vacuum state $|0,...,0\rangle$, 
\begin{equation}
    \mathcal{H}_V = \text{Span} \left\{|i_0^j,...,i_{d-1}^j\rangle \ | \  i_0^j +...+ i_{d-1}^j \neq 1 \right \}
\end{equation}
To recover $\mathcal{H}_S$, we can quotient the qubit Hilbert space $\mathcal{H}_Q$ by $\mathcal{H}_V$ and since $\mathcal{H}_Q = \mathcal{H}_S \oplus \mathcal{H}_V$, the quotient $\mathcal{H}_Q/\mathcal{H}_V$ is isomorphic to $\mathcal{H}_S$, which can be concretely represented as follows,\cite{roman2005advanced}
\begin{equation}
     \mathcal{H}_S = \text{Span}\left \{ |i_0^j,...,i_{d-1}^j\rangle \ | \  i_0^j + ... + i^j_{d-1} = 1 \right \}
\end{equation}
While we could work with $\mathcal{H}_Q$, projecting down to $\mathcal{H}_S$, yields a tremendous simplification of the qubit Hamiltonian. The projected kinetic energy operator is given by (see Supplementary Materials Sec. C for details), 
\begin{equation}
    l_j^2 = - \frac{1}{2}\sum_{m_j}^{d-1} (m_j - l)^2 \sigma_{m_j, j}^{z} + \frac{(2n^2 + 1)n}{6}
\end{equation}
Here, we have also imposed $n=l+1$. For the potential operators, we again need to encode the increment and decrement operators. Then, we can project down to $\mathcal{H}_S$ as we did for the kinetic energy operator. Using this approach, the interaction operators $x_ix_j$ are given by (see Supplementary Materials Sec. D for details and representations of other interaction operators), 
\begin{gather}
    x_ix_j = \frac{1}{4}\sum_{m_i,m_j}^{d} \sigma_{m_i+1,i}^{+} \sigma_{m_i,i}^{-} \sigma_{m_j+1,j}^{+} \sigma_{m_j,j}^{-} \nonumber \\ + \sigma_{m_i+1,i}^{+} \sigma_{m_i,i}^{-} \sigma_{m_j,j}^{+} \sigma_{m_j+1,j}^{-} + h.c.
\end{gather}
Unlike the binary encoding, this mapping results in constant interaction operator Pauli weights and the kinetic energy operator maps to a linear combination of on-site spin operators (no interaction terms). As discussed above, this is a consequence of using a larger number of encoding qubits and then projecting down to the subspace corresponding to the physical states of the rotor lattice. It is also instructive to note here that this encoding relies on initializing the system in the subspace $\mathcal{H}_S$. Then, evolving the state unitarily would necessarily restrict our system to the physical subspace. 

\subsection{Generalized Superfast Encoding}

As only nearest-neighbor interactions are involved in the Bosonic mapping, we can make a connection here between this system and a fermionic system with nearest-neighbor interactions. The equivalent Hamiltonian in fermionic space is
\begin{gather}
    H=C- h_0 + h_1 \\
h_0 = \sum_{n=0}^N \sum_{i=nd}^{(n+1)d}m_i^2 a_i^{\dagger}a_i \\ 
h_1 = \frac{g}{4}\sum_{n=0}^{N-1}\sum_{i=(n+1)d}^{(n+2)d-1}\sum_{k=(n)d}^{(n+1)d-1}\big(3 a_i^{\dagger} a_{i+1} a_k^{\dagger} a_{k+1} \nonumber \\ + 3 a_{i+1}^{\dagger} a_{i} a_{k+1}^{\dagger} a_{k} - a_i^{\dagger} a_{i+1} a_{k+1}^{\dagger} a_{k} \nonumber \\ -a_{i+1}^{\dagger} a_{i} a_k^{\dagger} a_{k+1}\big),
\end{gather}
where $C$ is a constant that depends on $N$ and $d$. 

This allows us to take advantage of a specific error detecting/correcting code; the Generalized Superfast Encoding (GSE).\cite{Setia_GSE} Here, the interaction graph includes only nearest neighbors. However, the Hamiltonian has the additional symmetry that the parity of $Z_{nd} ...Z_{nd+d-1}, n=0,N-1$ is even as all allowed states only have one spin down as shown in FIG. \ref{sparse_map}. Note that here we have used the notation: $\sigma^\alpha = \alpha$ for $\alpha \in \{X,Y,Z\}$, as we will do throughout this subsection. The interaction graph that will inherit this symmetry is a loop for each rotor's basis states. The interaction graph matrix representation for a distance-$d$ code would be 
\begin{equation}
    \begin{array}{cccccc}
         0 & d & 0 & ... & 0      &  d \\
         d & 0 & d & 0   &...     &0 \\
         0 & d & 0 &  \ddots   & \vdots & \vdots \\
    \vdots & 0 & \ddots & &d &0 \\
         0 & ... &  0 & d& 0 & d \\
         d & 0 & ... & 0 & d & 0
    \end{array}
\end{equation}
for each rotor basis. By judiciously choosing the local Majoranas,\cite{Setia_GSE} a tail-biting quantum convolutional code\cite{Forney_2007} is seen with a constant stabilizer weight of 6. This should allow error mitigation and possibly error correction for this problem on quantum hardware. It may also be possible to combine the inherent structure of the trotter evolutions of the encoded Hamiltonian with other error-correcting codes.\cite{Chen2025} As an example, the distance-3 (single qubit error correction) code has a two-node sliding description with logical $Z_l, Z_{l+1}$ of $Z_{dl}Z_{dl+1}Z_{dl+2}, Z_{dl+3}Z_{dl+4}Z_{dl+5}$ and stabilizers $Z_{dl}Y_{dl+1}X_{dl+2}Z_{dl+3}X_{dl+4}Y_{dl+5}$ and $ X_{dl}Z_{dl+1}Y_{dl+2}Y_{dl+3}Z_{dl+4}X_{dl+5}$. For example, for l=2, the distance 3 code would have stabilizers
\begin{equation}
    \begin{array}{|c|c|c|}
         ZYX  & ZXY & 0  \\
         XZY & YZX & 0  \\
         0 & ZYX & ZXY \\
    0 & XZY & YZX \\
         ZXY & 0 &  ZYX  \\
         YZX & 0 & XZY
    \end{array}
\end{equation} and logical Z and X of $ZZZ|ZZZ|ZZZ$ and $XXX|XXX|XXX$ respectively. 

\section{Quantum Algorithms} \label{algorithms}

\subsection{Quantum Phase Estimation} \label{sec:quantum_phase_estimation}

\begin{figure}[h]
\centering
\begin{quantikz}
    \lstick{\ket{0^r}} & \qwbundle{r} & \gate{U_{H}^{\otimes r}} & \ctrl{1} & \gate{F^{-1}_{r}} & \meter{} & \setwiretype{c} \rstick[1]{$\bar{\varphi}$} \\
    \lstick{\ket{\varphi}} &\qwbundle{k} && \gate{U} &&& \rstick[1]{\ket{\varphi}}
\end{quantikz}
\caption{Circuit diagram for the quantum phase estimation algorithm applied to an eigenstate of $U$. $U_{H}^{\otimes r}$ is the $r$ qubit Hadamard transform. The controlled-$U$ operation depicts a multiplicity-controlled-$U$ in this context and $F_{r}^{-1}$ is the $r$-qubit inverse quantum Fourier transform. $\bar{\varphi}$ is the $r$-bit approximation of the phase $\varphi$.} 
\label{qpe_circuit}
\end{figure}
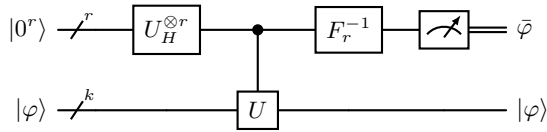

Quantum phase estimation (QPE) is a polynomial scaling quantum algorithm that can be used for extracting the phase $\varphi$ associated with the eigenvalue $e^{i 2 \pi \varphi}$ of a unitary operator $U$, given an eigenstate of $U$.\cite{mande2023tightboundsquantumphase, nielsenchuang} If the unitary in question is the time-evolution operator $e^{iHt}$, this can be used to estimate the energy of a given eigenstate.\cite{Setia_GSE} Fig. \ref{qpe_circuit} shows the circuit diagram for the algorithm. First, each of the $r$ qubits in the ancilla are Hadamard transformed. Then, a multiplicity-controlled $U$ is applied to the full state followed by an inverse Fourier transform which can be implemented with a polynomial gate cost. Finally, the ancilla encoding the $r$-bit binary representation of the phase $\bar{\varphi}$ associated with the input state $|\varphi\rangle$ is measured. The size of the ancilla $r$ is picked based on the desired accuracy of the phase $\varphi$. 

Note that to compute the spectrum of a bounded Hamiltonian $H$ using QPE, we must shift and scale the spectral radius so that all the eigenvalues $\tilde{E_i}$ of the resulting Hamiltonian $\tilde{H}$ can be uniquely mapped to a phase $\varphi \in [0,1)$.\cite{mande2023tightboundsquantumphase} This can be achieved by using the Hamiltonian, 
\begin{equation} \label{shift_scale_QPE_spectrum}
    \tilde{H} = \frac{1}{E_{max}^{c}-E_{min}^{c}} (H - E_{min}^{c}) 
\end{equation}
where $E_{max}^{c}$ and $E_{min}^{c}$ are the maximum and minimum classical energy bounds for the system. For a coplanar chain, the maximum classical energy is $E_{max}^{c} = Nn^2 + 2(N-1)g$, and the minimum classical energy is, $E_{min}^{c} = -2(N-1)g$. The multiplicity-controlled $U$ requires the application of $U^s$ for $0 \leq s \leq r$ to the state $|\varphi\rangle$. To implement these gates, we need to use an appropriate factorization technique for the real time propagator.\cite{higher_order_trotter, suzuki} In our computations, we find that the $4$th order Trotter approximation is sufficient to converge to the correct ground state energies with both encodings. 

Given a state with non-zero overlap with the ground state, we can also use QPE to prepare the exact ground state via repeated applications of the same circuit. For this we do not need to rescale the Hamiltonian if the ground state energy is known or if the approximate starting state has a sufficiently large overlap with the ground state. However, shifting the ground state energy (if it is known) so that it can be represented exactly using a small number of ancillary qubits available is useful for optimizing phase estimation simulations. The number of iterations of QPE required to project out the exact ground state is polynomial in the overlap between the trial state and the ground state.\cite{Setia_GSE} Consequently, QPE can also be used with approximate ground states to estimate the ground state properties such as the energy and the correlation function. Numerous techniques have been proposed in the literature for preparing the QPE trial state for many-body problems such as density matrix renormalization group (DMRG) and variational quantum eigensolver (VQE), a hybrid quantum-classical approach that we discuss in the next section.\cite{tilly2022variational, mande2023tightboundsquantumphase}

\subsection{Variational Quantum Eigensolver}

The large circuit depth requirement of QPE poses a significant challenge towards implementation on near-term devices.\cite{kanno2022resource, mande2023tightboundsquantumphase} Consequently, various near-term alternative approaches that require low-depth circuits have been proposed in the literature.\cite{TFIMVQE, sim2019expressibility} One such paradigm is VQE\cite{HVAVQE, heisenbergVQE} in which ground state expectation values are obtained by minimizing the energy corresponding to a unitary ansatz, 
\begin{equation}
    E(\vectheta) = \langle \psi_T|U(\vectheta)HU(\vectheta)|\psi_T\rangle
\end{equation}
Here, $\vectheta$ is the set of optimization parameters, $U(\theta)$ is a low-depth ansatz circuit that can be implemented on a near-term device and $|\psi_T\rangle$ is an easy-to-prepare state that has a non-zero overlap with the ground state. VQE proceeds in a hybrid manner, the optimization descent is performed classically while the expectation value is evaluated using a quantum device.\cite{HVAVQE}

\begin{figure*}[!t]
\centering
\begin{quantikz}
     & \gate{R_x\left(\theta_1^1 \right)} & \gate{R_y\left(\theta_2^1 \right)} & \gate{R_z\left(\theta_3^1 \right)} & \ctrl{2} && \ctrl{2} & \gate{R_x\left(\theta_4^1 \right)} & \gate{R_y\left(\theta_5^1 \right)} & \gate{R_z\left(\theta_6^1 \right)} & \\ 
     & \gate{R_x\left(\theta_1^2 \right)} & \gate{R_y\left(\theta_2^2 \right)} & \gate{R_z\left(\theta_3^2 \right)} & \ctrl{1} & \ctrl{1} && \gate{R_x\left(\theta_4^2 \right)} & \gate{R_y\left(\theta_5^2 \right)} & \gate{R_z\left(\theta_6^2 \right)} & \\ 
     & \gate{R_x\left(\theta_1^3 \right)} & \gate{R_y\left(\theta_2^3 \right)} & \gate{R_z\left(\theta_3^3 \right)} & \targ{} & \ctrl{1} & \ctrl{1} & \gate{R_x\left(\theta_4^3 \right)} & \gate{R_y\left(\theta_5^3 \right)} & \gate{R_z\left(\theta_6^3 \right)} & \\ 
     & \gate{R_x\left(\theta_1^4 \right)} & \gate{R_y\left(\theta_2^4 \right)} & \gate{R_z\left(\theta_3^4 \right)} && \targ{} & \targ{} & \gate{R_x\left(\theta_4^4 \right)} & \gate{R_y\left(\theta_5^4 \right)} & \gate{R_z\left(\theta_6^4 \right)} &
\end{quantikz}
\caption{Circuit diagram showing three layers of the full-entanglement $3$-local VQE ansatz with $CCX$ entangling gates.} 
\label{vqe_ansatz}
\end{figure*}
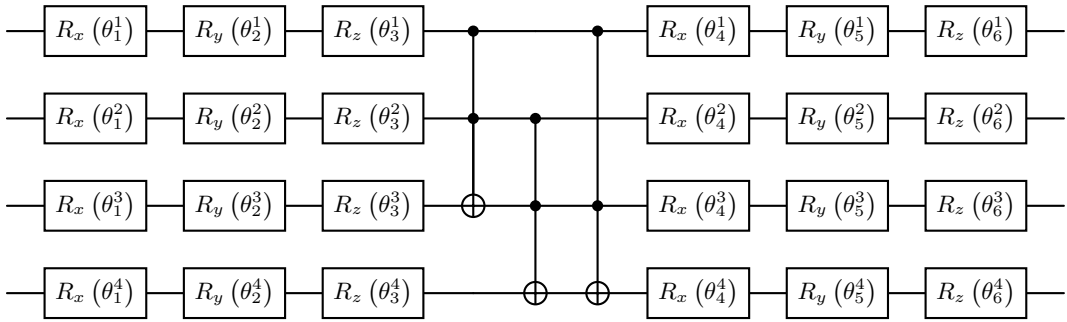

This can be done efficiently since the Hamiltonian can be decomposed into a sum of Pauli strings, each of which correspond to rotation gates. The expectation values of all of the required Pauli strings can be computed individually via repeated measurements and the resulting means summed to obtain the energy estimates. Since the number of Pauli strings in our Hamiltonian scales polynomially with the number of sites $N$, this approach's complexity relies on the efficiency of the optimization technique and the ansatz $U(\vectheta)$. Once an optimized set of variables $\vectheta$ have been obtained by minimizing the energy, ground state expectation values can be computed similarly by expressing the relevant operators as sums of Pauli strings. 

To explore the utility of VQE for our qubit encoding schemes, we use an fully entangled $N$-local ansatz\cite{kandala2017hardware} for both encodings. This is constructed by alternating layers of rotation gate blocks parameterized by optimization parameters ${\theta}$, and constant entanglement blocks. Figure $\ref{vqe_ansatz}$ shows three layers of the $3$-local circuit consisting of $CCX$ entangling gates that we use for our analysis. Each rotor is represented using two sequential qubits in this circuit. Moreover, the fully entangled nature of the middle block implies that the $CCX$ gate connects each qubit to every other qubit in the circuit. For our simulations, we will use a variant of this ansatz with $16$ blocks, each of which consists of one rotation layer and one entanglement layer. Additionally, a final rotation layer is also included in the ansatz as in Fig. \ref{vqe_ansatz}. 

\section{Computational Results \& Discussion} \label{section_3}
\subsection{Sparse Diagonalization}
\begin{figure}[h]
    \includegraphics[width=\linewidth, left]{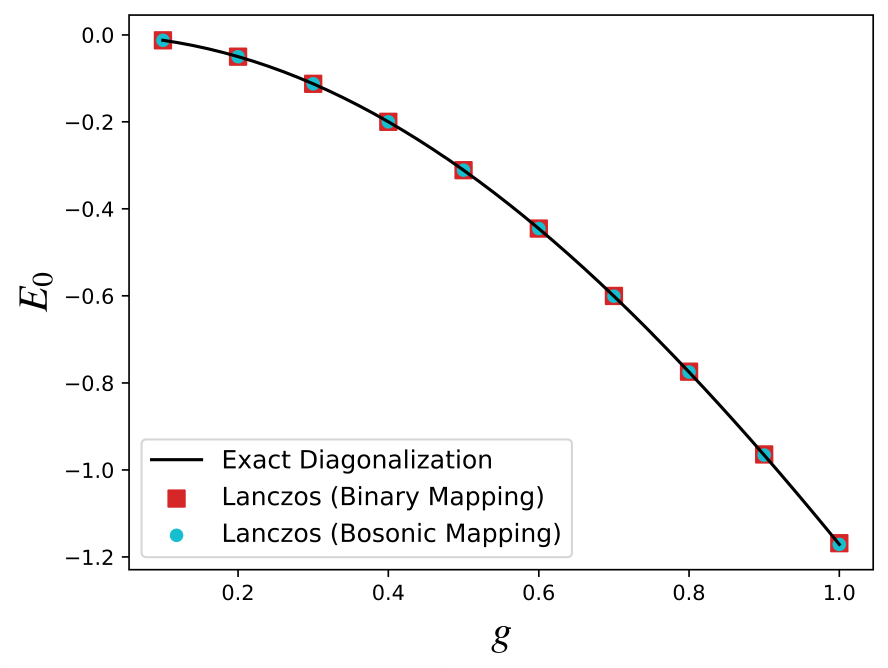}
    \caption{Ground state energy for $N=3$ computed using Lanczos diagonalization in the binary as well as the bosonic qubit representations. Exact results computed using the rotor Hamiltonian are exhibited for comparison.}
    \label{sparse_diag_results}
\end{figure}

To assess the validity of the qubit encodings discussed in Sec. \ref{binary_encoding} and Sec. \ref{bosonic_encoding}, we compute the ground state energy using Lanczos diagonalization for a coplanar chain with $N=3$.\cite{lehoucq1998arpack, lanczos1950iteration} Fig. \ref{sparse_diag_results} shows the results for both mappings for $g \in [0.1, 1.0]$. Exact diagonalization results calculated using the rotor momentum representation given in Eq. \eqref{rotor_momentum_representation_coplanar} are also plotted for comparison. 

For the binary mapping, we used $n=4$ and $l=3$. For the bosonic mapping, we used $n=3$ with $l=2$. In all of these cases, the mapped Hamiltonians yield the same ground state energy as exact diagonalization with the rotor representation. 

Note that the bosonic mapping results in a large Hilbert space dimension even for small systems. For $N=3$, the Hilbert space dimension with $n=4$ and $l=3$ is $2^{24}$. Consequently, we use linear operators for the sparse diagonalization simulations in Fig. \ref{sparse_diag_results} which can be constructed using the action of the Hamiltonian on an arbitrary vector in the lexicographically ordered computational qubit basis (see Supplementary Materials Sec. E for details). To simulate the binary mapping, we use the sparse qubit operator implementation provided by OpenFermion and Tangelo.\cite{openFermionRef, tangeloRef}

\subsection{Phase Estimation Trotter Convergence}
\begin{figure}[t]
    \includegraphics[width=\linewidth, left]{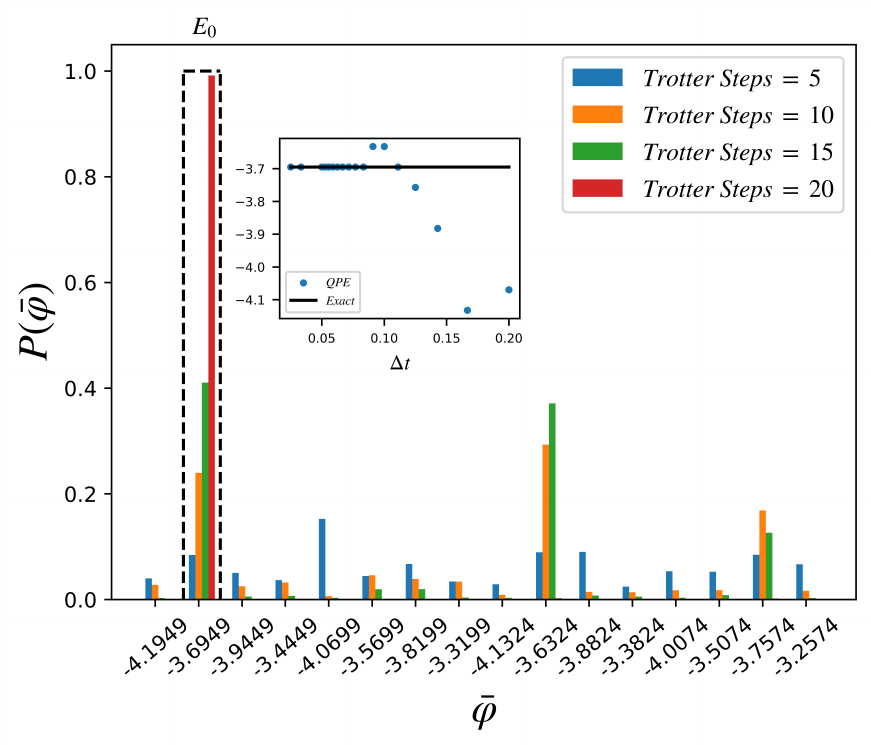}
    \caption{Probability of measuring any of the $16$ possible values of the phase $\bar{\varphi} = E_0$ associated with $U = e^{-i2\pi H}$ for $g=2.0$ with the binary mapping representation for different numbers of Trotter steps. The left panel corresponds to $N=2$ and the right panel is associated with $N=3$. The insets show convergence with $\Delta t$, the reciprocal Trotter steps. The exact $E_0$ value is indicated for reference.}
    \label{Trotter_convergence_histogram}
\end{figure}

Quantum phase estimation involves a multiplicity-controlled $U$ gate where $U = e^{itH}$. This needs to be evaluated using a factorization technique for a many-body system. For this work, we use a $4$th order Trotter factorization throughout. To investigate the minimum required number of Trotter steps ($P$) needed for convergence, we perform phase estimation simulations with $P \in \{5,10,15,20\}$. For this analysis, we shift the spectrum so that the ground state energy is $0.5$, 
\begin{equation}
    H \rightarrow H - (E_0 - 0.5) 
\end{equation}
which allows us to simulate the system with a small register. Fig. \ref{Trotter_convergence_histogram} shows the convergence of the phase estimation simulation using the binary mapping to the exact ground state energy for $N=2$ and $N=3$. For both values of $N$ we used $g=2.0$, $n=4$ and $l=3$ which yields $k=3$ and $d=8$. The bar charts show the probability of measuring all $16$ possible energy values given a $4$ qubit register. The exact $E_0$ is also indicated for reference. We find that $P = 20$ Trotter steps is sufficient for convergence in both cases. The insets show convergence in $\Delta t = 1/P$. For all of the simulations shown in Fig. \ref{Trotter_convergence_histogram}, the exact ground state was used as the input state $|\varphi\rangle$. We used \texttt{Tangelo} \cite{tangeloRef} with the \texttt{Qulacs} \cite{qulacsref} backend for this calculation.

\begin{figure}[b]
    \centering
    \includegraphics[width=\linewidth]{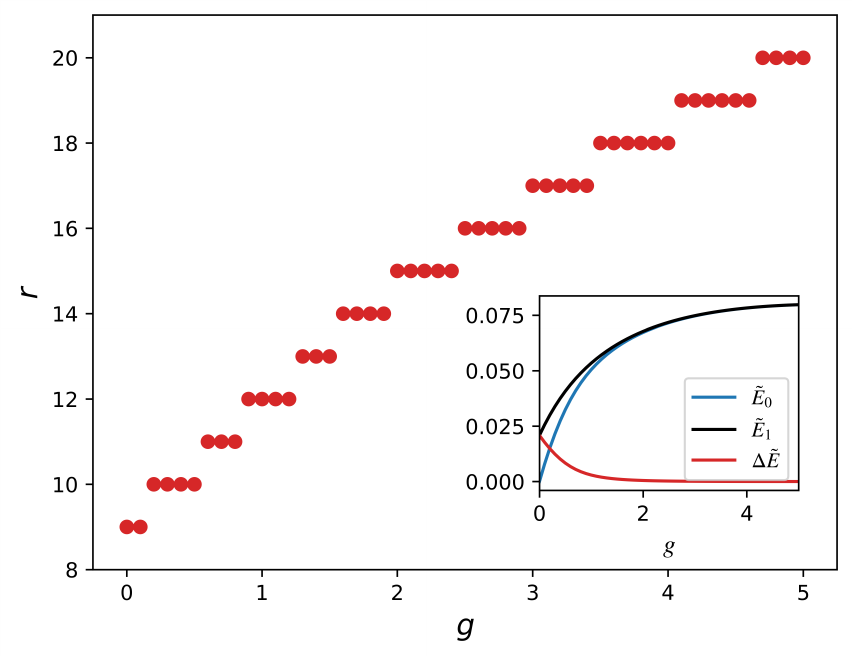}
    \caption{The lower bound on the QPE ancilla size for a coplanar chain of $3$ rotors for $g \in (0.0, 5.0]$, $n=4$ and $l=3$. The inset shows the first energy gap of the scaled Hamiltonian.}
    \label{reqd_ancilla_QPE_rotors}
\end{figure}

\subsection{Phase Estimation Register Requirements}
As discussed above, to compute the eigenvalues of a Hamiltonian in general, we need to rescale and shift the spectrum (see Eq. \eqref{shift_scale_QPE_spectrum}) so that all of the eigenvalues $\tilde{E}_i$ of the resulting Hamiltonian $\tilde{H}$ can be uniquely mapped to phases $\varphi \in [0.0, 1.0)$. This places a lower bound on the number of ancilla qubits required for QPE. For ground state calculations, note that the energy gap that needs to be resolved is $\Delta \tilde{E}_1 = \tilde{E}_1 - \tilde{E}_0$. 

If we use a $t$-bit approximation to measure $\tilde{E}_1$ and $\tilde{E}_0$, the precision error in each measurement is bounded above by $1/2^{t+1}$. Therefore, to resolve the two values separately, the energy gap $\Delta \tilde{E}_1$ must be larger than $2/2^{t+1} = 1/2^t$. This implies the following constraint on the number of bits needed to represent $\Delta \tilde{E}_1$, 
\begin{gather}
    \frac{1}{2^t} < \Delta \tilde{E}_1 \Rightarrow t > \log_2\left(\frac{1}{\Delta \tilde{E}}\right) \\ \Rightarrow t > \biggl \lceil \log_2\left(\frac{1}{\Delta \tilde{E}}\right) \biggr \rceil \Rightarrow t = \biggl \lceil \log_2\left(\frac{1}{\Delta \tilde{E}}\right) \biggr \rceil + 1
\end{gather}
In QPE, to obtain an estimate of the phase accurate to $t$ bits with a success probability of $1-\epsilon$, we need to use $r = t+s$ bits where $s$ is related to $\epsilon$ as follows,\cite{nielsenchuang}
\begin{equation}
    s = \lceil \log_2(2 + 1/2\epsilon) \rceil
\end{equation}
For $\epsilon = 1/12$, we get $s = 3$. This implies a QPE register bound on $r$, 
\begin{equation}
    r = \biggl \lceil \log_2\left(\frac{1}{\Delta \tilde{E}}\right) \biggr \rceil + 4
\end{equation}

\begin{figure*}[t]
    \includegraphics[width=\textwidth]{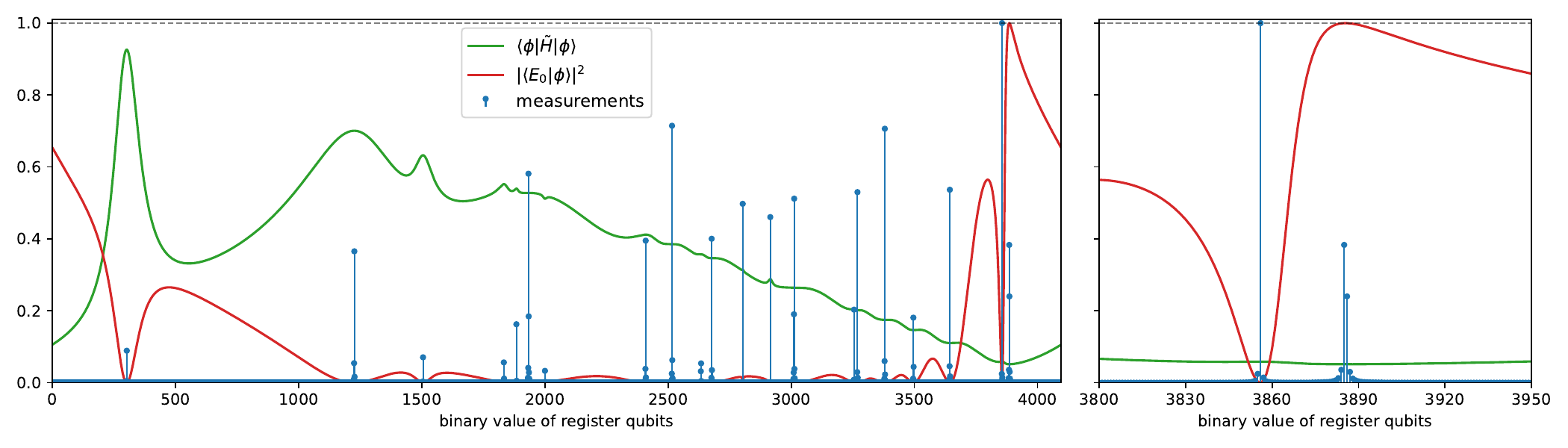}
    \caption{
        Results of the QPE simulations for a system of $ N = 2 $ rotors
        with an interaction strength of $ g = 1.5 $,
        using $ r = 12 $ ancilla qubits.
        After the simulation,
        we project the $ r $ ancilla qubits against an $ r $-digit binary string,
        leaving us with a $ k $-qubit statevector $ | \phi \rangle $.
        The horizontal axis shows the digital representation of the binary string
        of the ancilla qubits.
        The blue stems represent the counts of the post-simulation measurements,
        normalized such that the largest stem has a value of unity.
        The orange line shows $ | \langle E_0 | \phi \rangle |^2 $,
        the overlap of the projected state with the true ground state,
        the latter of which is found using exact diagonalization.
        The green line shows the expectation value of the rescaled Hamiltonian
        for the projected state $ \langle \phi | \tilde{H} | \phi \rangle $.
        The left plot shows the full spectrum of measurements,
        while the right plot focuses on the largest peak from the right of the spectrum,
        which corresponds to the ground state.
    }
    \label{fig:qpe_simulation_2rotor_12ancilla_g150}
\end{figure*}

\begin{figure*}[t]
    \includegraphics[width=\textwidth]{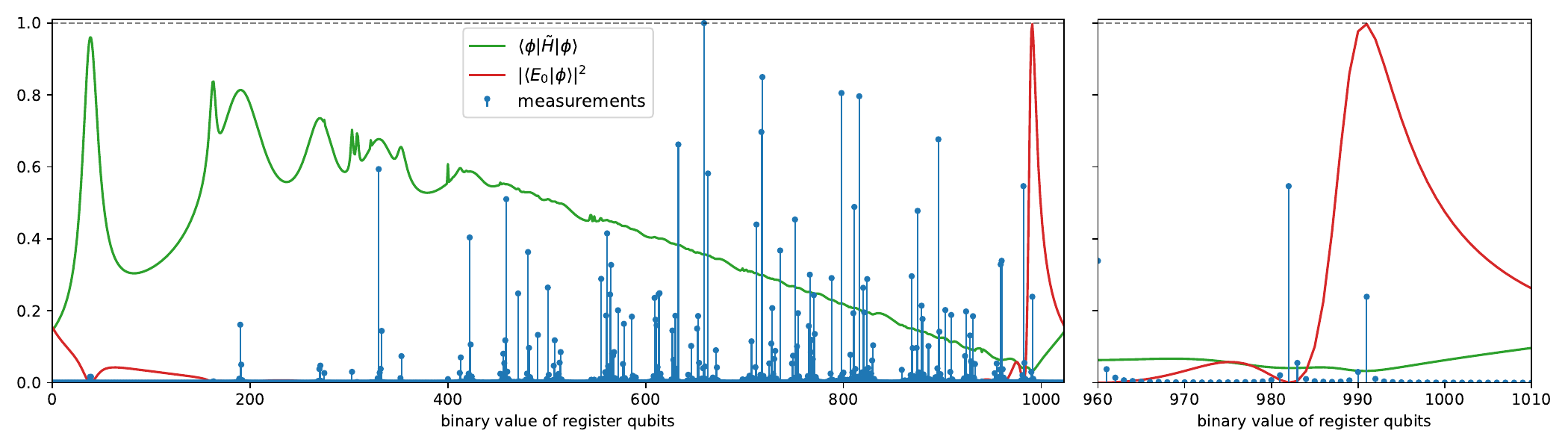}
    \caption{
        Results of the QPE simulations for a system of $ N = 3 $ rotors
        with an interaction strength of $ g = 0.5 $,
        using $ r = 10 $ ancilla qubits.
        The details of the plot match those given in the caption
        of Fig.~\ref{fig:qpe_simulation_2rotor_12ancilla_g150}.
    }
    \label{fig:qpe_simulation_3rotor_10ancilla_g050}
\end{figure*}

Fig. \ref{reqd_ancilla_QPE_rotors} displays the energy gap and the minimum required QPE ancilla size for a coplanar chain of $N=2$ and $N=3$ rotors for $g \in (0.0, 5.0]$. The gap was computed using exact diagonalization of the scaled planar rotor chain Hamiltonian in the momentum representation. The minimum and maximum scaling energies were set to the classical lower and upper bounds as discussed in Eq. \eqref{shift_scale_QPE_spectrum}. For $g \in (0.0, 1.0]$, the largest required ancilla size is $12$ qubits and the smallest is $9$ qubits. For a $3$ rotor chain with $n=4$ and $l=3$, we need $8$ qubits to encode the state. This implies that for a coplanar chain of $3$ rotors, QPE would require $\sim 20$ logical qubits. 

\subsection{Phase Estimation Simulations}

\begin{figure*}[t]
    \includegraphics[width=\textwidth]{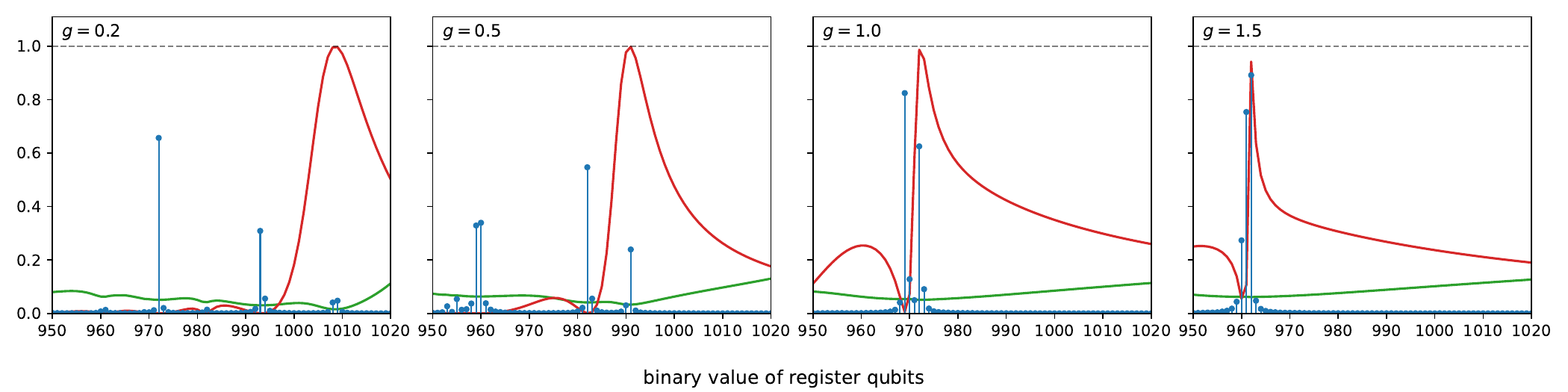}
    \caption{
        Results of the QPE simulations for a system of $ N = 3 $ rotors
        using $ r = 10 $ ancilla qubits,
        for four values of $ g $.
        The details of the plot match those given in the caption
        of Fig.~\ref{fig:qpe_simulation_2rotor_12ancilla_g150}.
    }
\label{fig:gaps_n_rotors_3_registers_10_classical_hadamard}
\end{figure*}

In this section, we discuss our noiseless QPE simulation results targeting dipolar chains of $N=2$ and $N=3$ rotors with a rescaled Hamiltonian following Eq.~(\ref{shift_scale_QPE_spectrum}). 
Based on the results of Fig. \ref{Trotter_convergence_histogram},
we used $ P = 20 $ Trotter steps to ensure convergence.
For $ N = 2 $ rotors, $r=12$
and for $ N = 3 $ rotors $r$ was set to $10$.
Results corresponding to $g=1.5$ for $N=2$ and $g=0.5$ for $N=3$ are shown in Figs, \ref{fig:qpe_simulation_2rotor_12ancilla_g150} and \ref{fig:qpe_simulation_3rotor_10ancilla_g050} respectively. 
For the input state we used the uniform superposition state
$ | \varphi \rangle = U_{H}^{\otimes Nk} |0\rangle^{\otimes Nk} $ where $U_H$ represents the single qubit Hadamard transform.
This initial guess state
has sufficient overlap with many of the eigenstates,
including the ground state,
while also being general enough to
avoid requiring prior knowledge of those eigenstates. These QPE simulations were performed using \texttt{kettle},\cite{kettle_citation}
our in-house quantum circuit simulator. 

In Fig.~\ref{fig:qpe_simulation_2rotor_12ancilla_g150},
we show the QPE simulation results for the case $ (N, r, g) = (2, 12, 1.5) $.
At each peak,
the projected statevector closely approximates an eigenstate.
In particular,
at the rightmost significant peak,
the projected statevector approximates the ground state.
As an indication of this,
we can see that the projected statevectors
around the rightmost peak have an overlap with the true ground state (shown by the orange line)
of nearly unity.
We also show the expectation value of the rescaled Hamiltonian corresponding to the QPE projected state (exhibited using the green line) for reference. It achieves its minimum at the rightmost significant peak (as expected for a good approximation of the ground state). 

The value of the rescaled energy of the QPE projected state
around the first peak (from the right) is $ 5.140673 \times 10^{-2} $,
which closely approximates the ground state eigenvalue of $ 5.140588 \times 10^{-2} $. 
found through exact diagonalization.
As we move from right to left in Fig.~\ref{fig:qpe_simulation_2rotor_12ancilla_g150},
the peaks correspond to eigenstates of increasing energy.
Using this approach, we can recover not only an approximation to the ground state,
but also approximations to all eigenstates that
have sufficient overlap with the initial input state.

Fig.~\ref{fig:qpe_simulation_3rotor_10ancilla_g050},
shows the QPE simulation results for the case $ (N, r, g) = (3, 10, 0.5) $.
With fewer ancilla qubits, 
the approximations of the eigenstates are not as accurate as
for the aforementioned case of $ (N, r, g) = (2, 12, 1.5) $ since our QPE success probability decreases as we decrease the register size.
For example,
the expectation values have fewer significant digits of accuracy;
at the rightmost peak the expectation value is $ 3.25421 \times 10^{-2} $,
compared to the value of about $ 3.24766 \times 10^{-2} $
found through exact diagonalization. However, a good approximation to the ground state is still recovered as exhibited by the overlap $|\langle E_0|\phi\rangle|^2$ achieving a value of close to $1$ at the rightmost significant peak.   

In Fig.~\ref{fig:gaps_n_rotors_3_registers_10_classical_hadamard},
we show QPE simulation results for a system of $ N = 3 $
simulated with $ r = 10 $ ancilla qubits,
for four different values of $ g $.
The peak associated with the ground state
is identifiable by the position of the maximum of the orange curve,
which again represents 
the overlap of the projected state with the true ground state.
For the smallest value chosen, $ g = 0.2 $,
the peaks corresponding to the ground state and first excited state
are still clearly distinguishable.
As the value of $ g $ increases,
the peaks for the ground state and first excited state move closer.
In the case where $ g = 1.5 $,
the rescaled ground and first excited state energies are close enough
that the two peaks are no longer resolvable. This is because at $g=1.5$, the minimum bound on the QPE register is $t = -\lceil \log_2(\Delta \tilde{E}) \rceil + 4 = 13$ and our simulation used $r=10$. 

It is also instructive to note that as $g$ gets smaller, the height of the peak corresponding to the maximum ground state overlap decreases. This implies that the probability of projecting out the ground state in any single QPE run decreases. This is because as $g$ gets closer to $0$, the Hamiltonian is dominated by the kinetic energy operator which implies that the ground state in that regime is a product state. Since our initial state was a (normalized) uniform superposition of all product states, the overlap of any single product state with the initial state is $1/2^{Nk}$. Contrasting that with the case of large $g$, in the limit of $g \rightarrow \infty$, the ground state in the momentum representation is a uniform superposition all states $|m_1, m_2,m_3\rangle$ such that $M = m_1 + m_2 + m_3$ is even.\cite{estevao_pimc} Therefore, as $g$ gets larger, the overlap of the initial state and the ground state approaches $N_M/2^{Nk} > 1/2^{Nk}$ where $N_M$ is the number of states with $M$ even, resulting in the increasing peak sizes. 

\begin{figure*}[t]
    \includegraphics[width=0.47\textwidth]{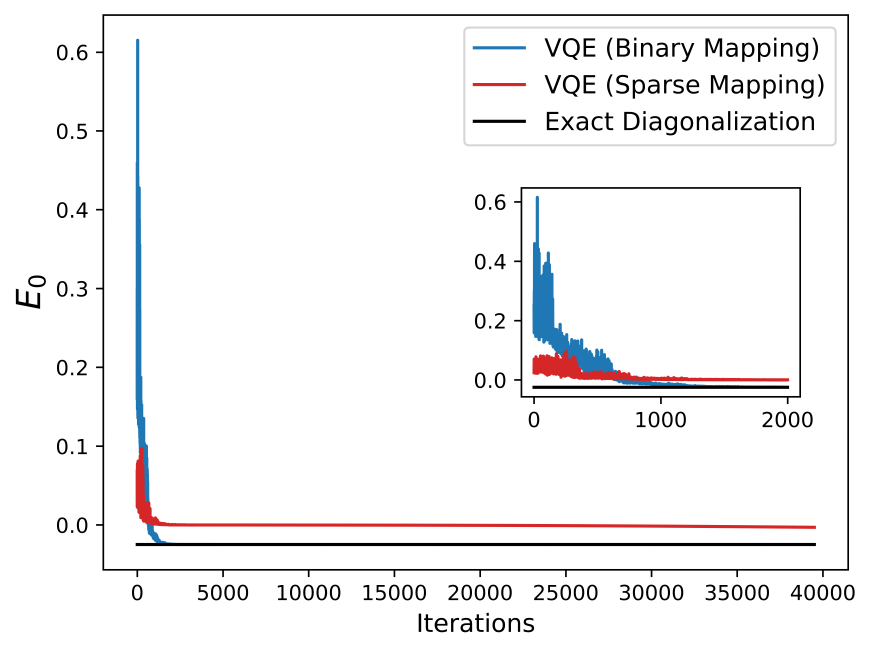}
    \includegraphics[width=0.49\textwidth]{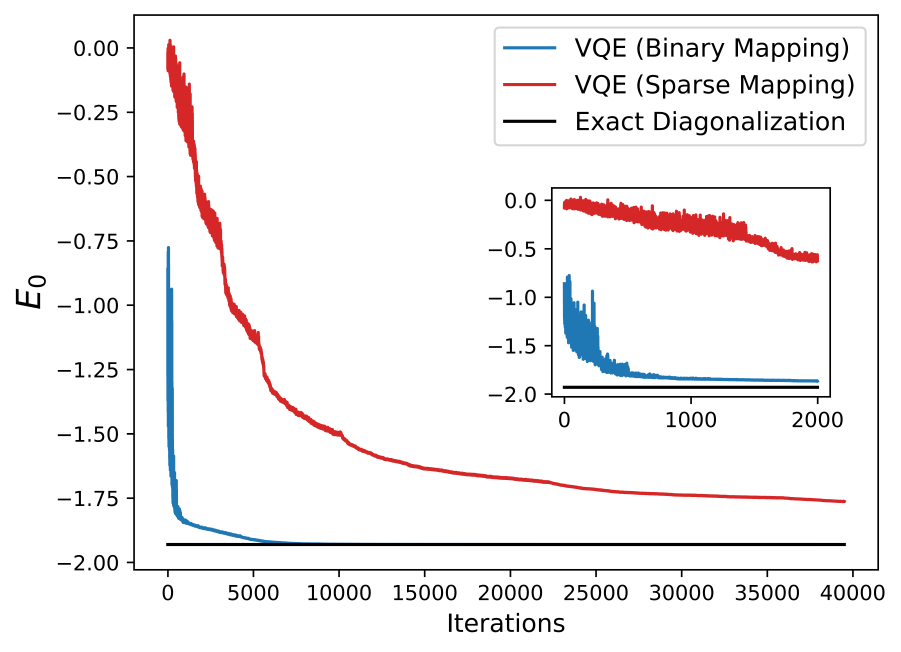}
    \caption{VQE results for a chain of $2$ rotors with both encoding schemes. For these calculations, we used $n=2$ and $l=1$. The left panel corresponds to $g=0.2$ and the right panel exhibits the results for $g=2.0$. Insets show the first $2000$ iterations.}
    \label{VQE_convergence}
\end{figure*}

\subsection{Variational Simulations}

We also perform VQE simulations to asses the feasibility of studying dipolar rotor lattices using such devices. We simulate a system of $N=2$ rotors with $l=1$ and $n=2$, representing each rotor using $2$ qubits for the binary encoding, and $4$ qubits in the case of the unary mapping (since $d=4$). In both cases, we use the fully entangled $3$-local ansatz structure exhibited in Fig. \ref{vqe_ansatz} with $16$ blocks. For all of our VQE simulations, we use the COBYLA optimizer\cite{powell1994direct} and set the initial state to $|m_1=0,m_2=0\rangle$ which maps to $|0101\rangle$ in the binary encoding and $|00010001\rangle$ in the unary encoding when $d=4$. 

The results for $g=0.2$ and $g=2.0$ are shown in Fig. \ref{VQE_convergence}. The binary mapping converges relatively quickly (less than $10^4$ iterations for large $g$ and less than $10^3$ iterations for small $g$). However, for the unary mapping, the simulation relaxes into a local minimum. This is likely because the state space explored with the bosonic mapping is much larger than with the binary mapping due to the presence of states in the qubit Hilbert space that do not correspond to any state in the physical rotor Hilbert space as discussed in Sec. \ref{bosonic_encoding}. This could potentially be resolved by using a particle-number preserving ansatz such as a unitary coupled-cluster (UCC) ansatz.\cite{yordanov2021qubit} This potential avenue for improvement will be explored seperately in conjunction with other approaches to characterize the behavior of various ansatz circuits for the dipolar rotor problem. 

\section{Conclusions \& Outlook} \label{conclusions_outlook}

In this work, we introduced two qubit encoding schemes for lattices of planar rotors. The first encoding relies on using the binary decomposition of the discrete momentum representation. The second mapping embeds the rotor Hilbert space into a larger space and recovers the physical degrees of freedom projectively. Focusing on a $1$ dimensional lattice (chain) with a nearest-neighbor dipolar coupling, we computationally verified these representations using sparse diagonalization. We also performed quantum phase estimation\cite{nielsenchuang} simulations for $N=2$ and $N=3$ to examine the resource requirements for simulating dipolar rotor lattices using quantum devices. 

Finally, since phase estimation has steep circuit depth and error-correction requirements,\cite{kanno2022resource, mande2023tightboundsquantumphase} we explored the utility of using variational approaches to compute ground state properties of our system. This was done with a fully entangled $3$-local ansatz.\cite{kandala2017hardware} While we found that this leads to the well-documented barren plateau problem\cite{HVAVQE} when the bosonic encoding is used, the binary mapping converges to the correct ground state efficiently. This suggests that the $3$-local ansatz could be used in a VQE paradigm to study the ground state of the dipolar planar rotor lattices using a quantum device. 

As discussed above, particle preserving-ansatz circuits (such as the UCC ansatz)\cite{yordanov2021qubit} would be good candidates for the bosonic mapping since they would prevent the simulation from traversing states that do not correspond to physical rotor states. Moreover, our VQE simulations were both noise-less. Adding in the effect of noise\cite{MPSVQE} as well as characterizing the efficiency of different ansatz architectures will be explored in future works. Another interesting future direction would be to use the qubit encodings developed here to develop better classical algorithms for simulating dipolar rotor lattices. 

In particular, current quantum Monte Carlo capabilities targeting planar rotor lattices experience critical slowing down near critical points.\cite{wolff1989critical} However, for spin lattices, efficient global move algorithms have been developed such as stochastic series expansion approaches which mitigate this loss of efficiency near critical points using global transitions in the $(L+1)$ dimensional configuration space.\cite{directed_loop_qmc, sse_tfim} Therefore, since these qubit encodings map an $L$ dimensional lattice of rotors to an $(L+1)$ dimensional lattice of spins, it would be very interesting to examine the utility of global move  quantum Monte Carlo algorithms for studying lattices of dipolar rotors.  

\section*{Supplementary Material}
See the supplementary materials for details pertaining to the qubit encoded Hamiltonians from Sec. \ref{theory} that were omitted from the main text for brevity. 

\section*{acknowledgments}
This research was supported by the Natural Sciences and Engineering Research Council (NSERC) of Canada (RGPIN-03725-2022), the Ontario Ministry of Research and Innovation (MRI), the Canada Research Chair program (950-231024), the Digital Research Alliance of Canada, and the Canada Foundation for Innovation (CFI) (project No. 35232).

\section*{Data Availability Statement}
The data that support the findings of this study are available
from the corresponding author upon reasonable request.

\section*{References}
\bibliography{refs}

\pagebreak
\clearpage
\onecolumngrid

{
\sffamily
\bfseries
\linespread{1.5}
\Large 
\noindent
Supplementary Material for: Qubit encodings for lattices of dipolar planar rotors
}

\setcounter{equation}{0}
\setcounter{figure}{0}
\setcounter{table}{0}
\setcounter{page}{1}
\setcounter{section}{0}
\makeatletter
\renewcommand{\theequation}{S\arabic{equation}}
\renewcommand{\bibnumfmt}[1]{[S#1]}
\renewcommand{\citenumfont}[1]{S#1}

\setlength{\parindent}{0pt}

\thispagestyle{empty}

\onecolumngrid

\vspace{1.0em}
Here, we provide additional details pertaining to the representation of the planar rotor Hamiltonian using a lattice of spin-1/2 particles. 

\subsection{Binary Encoding: Kinetic Energy Operator} \label{supp_sect_a}
The kinetic energy operator in the momentum representation can be expressed using the binary encoding as done in the main text, 
\begin{gather}
    l_{j}^2 = \sum_{m_j = -l}^{n} m_j^2 |m_j\rangle\langle m_j| =  \sum_{m_j = 0}^{d-1} (m_j - l)^2 |m_j\rangle \langle m_j| \nonumber \\ = \sum_{i_1^j, ..., i_k^j} \left(\sum_{r=1}^{k} 2^{k-r} i_{r}^j-l\right)^2 |i_1^j,...,i_k^j\rangle \langle i_1^j, ..., i_k^j| \label{qbit_rep_1_diagonal}
\end{gather}
Here, we have used $d = n+l+1$, and $k=\lceil \log_2(d) \rceil$ to represent the number of qubits. To construct qubit operators from this, we can expand the square of the eigenvalues of $l_j$ in Eq. \eqref{qbit_rep_1_diagonal} to get,
\begin{gather}
    l_{j}^2 = \sum_{i_1^{j}, ..., i_k^{j}} \sum_{r,s=1}^{k} 2^{2k-r-s} i_r^j i_s^j |i_1^j,...,i_k^j\rangle \langle i_1^j, ..., i_k^j| \nonumber \\ - 2l \sum_{i_1^{j}, ..., i_k^{j}}\sum_{r=1}^{k} 2^{k-r} i_r^j |i_1^j,...,i_k^j\rangle \langle i_1^j, ..., i_k^j| + l^2 \\ \Rightarrow l_{j}^2 = 2^{2k}\sum_{r,s} 2^{-r-s}\sum_{i_r^{j}, i_s^{j}} i_{r}^{j} i_{s}^{j} |i_r^{j}, i_s^{j} \rangle \langle i_r^{j}, i_s^{j}| - 2^{k+1}l \sum_{r} 2^{-r} \sum_{i_r^j} i_r^{j} |i_r^{j}\rangle \langle i_s^{j}| + l^2 
\end{gather}
Since $i_r,  i_s \in \{0,1\}$, we need only account for terms corresponding to $i_r, i_s = 1$, 
\begin{gather}
    l_j^2 = 2^{2k}\sum_{r,s} 2^{-r-s} |1_{r}^{j}, 1_{s}^{j}\rangle \langle 1_{r}^{j}, 1_{s}^{j}| -2^{k+1} l \sum_{r} 2^{-r} |1_{r}^j\rangle \langle 1_{r}^j| + l^2 
\end{gather}
Now, using the fact that $|1\rangle \langle 1| = \frac{1}{2} (I - \sigma_z)$, we get, 
\begin{gather}
    l_j^2 = 2^{2(k-1)} \sum_{r,s} 2^{-r-s} (I-\sigma_{r,j}^{z}) (I-\sigma_{s,j}^{z}) - 2^{k} l \sum_{r} 2^{-r}(I-\sigma_{r,j}^{z}) + l^2 
\end{gather}
Expanding out the products yields, 
\begin{gather}
    l_j^2 = 2^{2(k-1)} \sum_{r,s} 2^{-r-s} \sigma_{r,j}^{z} \sigma_{s,j}^{z} - 2^{2k-1}\sum_{s=1}^k 2^{-s}\sum_{r} 2^{-r} \sigma^{z}_{r,j} + 2^{k} l \sum_{r} 2^{-r} \sigma^{z}_{r,j} \nonumber \\ + 2^{2k-2} \left(\sum_{r=1}^{k} 2^{-r}\right)^2 - 2^{k}l \sum_{r=1}^{k} 2^{-r} + l^2 \label{momentum_pauli_z_rep}
\end{gather}
If we restrict $d$ to be a power of $2$ and collect terms linear in $\sigma^z$, we get,  
\begin{gather}
    -2^{k-1} (2^k - 1 - 2l) \sum_{r=1}^{k} 2^{-r} \sigma_{r,j}^{z} = -\frac{(n+l+1)(n-l)}{2} \sum_{r=1}^{k} 2^{-r} \sigma^{z}_{r,j}
\end{gather}
where we have used the following,
\begin{equation}
    \sum_{r=1}^{k} 2^{-r} = 1-2^{-k}, \ \ \ 2^{k} = n + l + 1
\end{equation}
For the constant term in Eq. \eqref{momentum_pauli_z_rep}, we similarly get, 
\begin{equation}
    \frac{1}{4} (2^{k} - (2l + 1))^2 = \frac{1}{4}(n-l)^2
\end{equation}
This finally gives the following for the kinetic energy operator, 
\begin{gather}
    l_{j}^{2} = \left(\frac{n+l+1}{2}\right)^2 \sum_{r,s} 2^{-r-s} \sigma_{r,j}^{z} \sigma_{s,j}^{z} - \frac{(n+l+1)(n-l)}{2} \sum_{r=1}^{k} 2^{-r} \sigma^{z}_{r,j} + \frac{1}{4}(n-l)^2 \label{single_rotor_ke_first_map}
\end{gather} 

\subsection{Binary Encoding: Position Operators} \label{supp_sect_b}
To construct the Pauli string representation of $x_j$ using the binary encoding, we start with the operator $x_j$ represented using the increment and decrement operators, 
\begin{gather}
    x_j = \frac{1}{2}(S_j^{+} + S_j^{-}) = \frac{1}{2}\sum_{r=1}^{k} \left(\sigma_{r,j}^{+} \prod_{s=r+1}^{k}\sigma_{s,j}^{-} + \sigma_{r,j}^{-} \prod_{s=r+1}^{k}\sigma_{s,j}^{+}\right) + \frac{1}{2}\prod_{s=1}^{k}\sigma_{s,j}^{-} + \frac{1}{2}\prod_{s=1}^{k}\sigma_{s,j}^{+} \label{x_j_first_eqn}
\end{gather}
Here, we have used the binary encoding of the $S_j^+$ and $S_j^-$ as discussed in the main text, 
\begin{gather}
    S^+_j = \sum_{m_j=0}^{d-1}|m_j+1\rangle \langle m_j|  =\sum_{r=1}^{k} \sigma_{r,j}^{+} \left(\prod_{s=r+1}^{k}\sigma_{s,j}^{-}\right) + \left(\prod_{s=1}^{k}\sigma_{s,j}^{-}\right) \label{binary_mapping_S_plus} 
\end{gather}
\begin{gather}
    S_{j}^{-} = \sum_{m_j=0}^{d-1} |m_j-1\rangle \langle m_j|  = \sum_{r=1}^{k} \sigma_{r,j}^{-} \left(\prod_{s=r+1}^{k}\sigma_{s,j}^{+}\right) + \left(\prod_{s=1}^{k}\sigma_{s,j}^{+}\right)
\end{gather}

For each term in the sum of Eq. \eqref{x_j_first_eqn} corresponding to $r$ where $1 \leq r \leq k$, we can also construct the associated Pauli string using the definitions of $\sigma_{r}^{\pm}$. For each such term $(x_{j})_r$, we get,
\begin{gather}
    (x_{j})_{r} = \frac{1}{2^{k-r+1}} \sigma_{r,j}^{x}\sum_{\gamma \in \{x,y\}^{k-r}} (-i)^{n_y} \prod_{s=r+1}^{k} \sigma_{s,j}^{\gamma(s)} \nonumber + \frac{i}{2^{k-r+1}} \sigma_{r,j}^{y}\sum_{\gamma \in \{x,y\}^{k-r}} (-i)^{n_y} \prod_{s=r+1}^{k} \sigma_{s,j}^{\gamma(s)} \nonumber \\ + \frac{1}{2^{k-r+1}} \sigma_{r,j}^{x}\sum_{\gamma \in \{x,y\}^{k-r}} i^{n_y} \prod_{s=r+1}^{k} \sigma_{s,j}^{\gamma(s)} - \frac{i}{2^{k-r+1}} \sigma_{r,j}^{y}\sum_{\gamma \in \{x,y\}^{k-r}} i^{n_y} \prod_{s=r+1}^{k} \sigma_{s,j}^{\gamma(s)}
\end{gather}
Here, we sum over all $2^{k-r}$ possible vectors $\gamma = (\gamma(r+1), ..., \gamma(k))$ with $\gamma(s) \in \{x,y\}$ for each term. $n_y$ is the number of indices $s \in \{r+1, s\}$ such that $\gamma(s) = y$. Adding the first and third terms in the above equation yields a cancellation of all terms corresponding to an odd $n_y$. Similarly, adding the second and the last terms results in a cancellation of all terms associated with an even $n_y$. This gives, 
\begin{gather}
    x_{j} = \frac{1}{2}\sum_{r=1}^{k} \frac{1}{2^{k-r}} \Biggl( \sum_{\gamma \ \text{even}} (-1)^{n_y/2} \sigma_{r,j}^{x} \prod_{s=r+1}^{k} \sigma_{r,j}^{\gamma(s)} + \sum_{\gamma \ \text{odd}} (-1)^{(n_y-1)/2} \sigma_{r,j}^{y}\prod_{s=r+1}^{k} \sigma_{s,j}^{\gamma(s)}\Biggr) \nonumber \\ + \frac{1}{2^k} \sum_{\gamma \ \text{even}} (-1)^{n_y/2} \prod_{s=1}^{k} \sigma_{s,k}^{\gamma(s)}
\end{gather}
where the last term comes from the operator strings corresponding to products of only $\sigma^{+}$ and products of only $\sigma^{-}$ terms in the definition of the increment and decrement operators. Note that in the above equation $\gamma \ \text{even}$ denotes index strings with an even number of elements equaling $y$ and vice versa for $\gamma \ \text{odd}$. For $y_j$, we similarly get, 
\begin{gather}
    y_j = \frac{1}{2}\sum_{r=1}^{k} \frac{1}{2^{k-r}} \Biggl( \sum_{\gamma \ \text{odd}} (-1)^{(n_y+1)/2} \sigma_{r,j}^{x} \prod_{s=r+1}^{k} \sigma_{r,j}^{\gamma(s)} + \sum_{\gamma \ \text{even}} (-1)^{n_y/2} \sigma_{r,j}^{y}\prod_{s=r+1}^{k} \sigma_{s,j}^{\gamma(s)}\Biggr) \nonumber \\ + \frac{1}{2^k} \sum_{\gamma \ \text{odd}} (-1)^{(n_y+1)/2} \prod_{s=1}^{k} \sigma_{s,k}^{\gamma(s)}
\end{gather}
Note that the matrix elements of $x_j$ are all real and the matrix elements of $y_j$ are all imaginary as expected from their respective definitions in the momentum representation.

\subsection{Bosonic Encoding: Projected Kinetic Energy Operator}
The kinetic energy operator in the Bosonic qubit  encoding is given by, 
\begin{equation}
    l_j^2 = \sum_{m_j=0}^{d-1} \frac{(m_j - l)^2}{2^d} (I - \sigma_{m_j,j}^{z}) \prod_{k \neq m_j}^{d-1}(I + \sigma_{k,j}^{z}) \label{bosonic_mapping_ke_1}
\end{equation}
As discussed in the main text, the qubit Hilbert space is $\mathcal{H}_Q = (\mathbb{C}^{2})^{\otimes d}$ with dimension $2^{d}$ while our single rotor system Hilbert space $\mathcal{H}_S$ has dimension $d$. The extra degrees of freedom correspond to the set $\mathcal{H}_V \subset \mathcal{H}_Q$ defined by the span of all product states with more than one qubit in state $|1\rangle$, and the vacuum state $|0,...,0\rangle$, 
\begin{equation}
    \mathcal{H}_V = \text{Span} \left\{|i_0^j,...,i_{d-1}^j\rangle \ | \  i_0^j +...+ i_{d-1}^j \neq 1 \right \}
\end{equation}
To recover $\mathcal{H}_S$, we quotient the qubit Hilbert space $\mathcal{H}_Q$ by $\mathcal{H}_V$ and since $\mathcal{H}_Q = \mathcal{H}_S \oplus \mathcal{H}_V$, the quotient $\mathcal{H}_Q/\mathcal{H}_V$ is isomorphic to $\mathcal{H}_S$, which can be concretely represented as follows, 
\begin{equation}
     \mathcal{H}_S = \text{Span}\left \{ |i_0^j,...,i_{d-1}^j\rangle \ | \  i_0^j + ... + i^j_{d-1} = 1 \right \}
\end{equation}
The kinetic energy operator, restricted to $\mathcal{H}_S$, can be obtained by examining the action of $l_j^2$ on $\mathcal{H}_S$, 
\begin{gather}
    l_j^2 |\psi\rangle 
    = \sum_{m_j=0}^{d-1}(m_j-l)^2 \frac{(I - \sigma_{m_j,j}^{z})}{2}  |\psi\rangle  
\end{gather}
where $|\psi\rangle$ is an arbitrary linear combination of product states in $\mathcal{H}_S$. Here, we have used the fact that for a product state $|\psi_r\rangle = |0^{r-1}\rangle \langle 0^{r-1}| \otimes |1\rangle \langle 1| \otimes |0^{d-r}\rangle \langle 0^{d-r}|  \in \mathcal{H}_S$, 
\begin{gather}
    \left(\frac{I - \sigma_{m_j,j}^{z}}{2}\right) \prod_{k\neq m_j}^{d-1} \left(\frac{I + \sigma_{k,j}^{z}}{2}\right) |\psi_r\rangle = \left(\frac{I - \sigma_{m_j,j}^{z}}{2}\right)|\psi_r\rangle
\end{gather} 
for all $0 \leq r \leq d-1$. Note that if we include states $|\psi\rangle \in  \mathcal{H}_V$, the above equality does not hold. To see this, consider any product state $|\psi\rangle \in \mathcal{H}_V$ such that the qubit corresponding to index $m_j$ is in state $|1\rangle$. Then, by definition of $\mathcal{H}_V$, there exists $r \neq m_j, \ r \in \{0,...,d-1\}$ such that $i_r^j = 1$. This implies, 
\begin{gather}
    \left(\frac{I - \sigma_{m_j,j}^{z}}{2}\right) \prod_{k\neq m_j}^{d-1} \left(\frac{I + \sigma_{k,j}^{z}}{2}\right) |\psi\rangle = \left(\frac{I - \sigma_{m_j,j}^{z}}{2}\right) |1\rangle \prod_{k\neq m_j}^{d-1} \left(\frac{I + \sigma_{k,j}^{z}}{2}\right) |0\rangle \left(\frac{I + \sigma_{k,j}^{z}}{2}\right) |1\rangle = 0 \nonumber 
\end{gather}
However, if we consider just the projector corresponding to $m_j$, we get, 
\begin{equation}
    \left(\frac{I - \sigma_{m_j,j}^{z}} {2}\right)|\psi\rangle = |\psi\rangle \neq 0 \nonumber
\end{equation}
Therefore, only if we restrict to $\mathcal{H}_{S}$ can we neglect terms corresponding to $|0_{k}\rangle \langle 0_{k}|$ in our Hamiltonian. Using this projection, our kinetic energy is linear in Pauli operators, 
\begin{gather}
    l_j^2 = - \frac{1}{2}\sum_{m_j=0}^{d-1} (m_j - l)^2 \sigma_{m_j, j}^{z} + \frac{n(n+1)(2n+1) + l(l+1)(2l+1)}{12} 
\end{gather}
where we have used Faulhaber's formula to compute the constant energy offset. Setting $n=l+1$, we get, 
\begin{equation}
    l_j^2 = - \frac{1}{2}\sum_{m_j=0}^{d-1} (m_j - l)^2 \sigma_{m_j, j}^{z} + \frac{(2n^2 + 1)n}{6}
\end{equation}

\subsection{Bosonic Encoding: Interaction Operators}

For the interaction operators in the bosonic encoding, we first construct the increment and decrement operators by mapping the projectors from the rotor momentum representation to the qubit representation. To this end, consider the projection operators $|m_j+1\rangle \langle m_j|$,
\begin{gather}
    |m_j+1\rangle \langle m_j| = |0^{m_j-1}\rangle \langle 0^{m_j-1}| \otimes |0\rangle \langle 1| \otimes |0 \rangle \langle 1| \otimes |0^{d - m_j - 1} \rangle \langle 0^{d - m_j - 1}| \nonumber \\ = \sigma_{m_j+1,j}^{+} \sigma_{m_j,j}^{-} \prod_{k \neq m_j,m_{j+1}}^{d-1}\left(\frac{I+\sigma_{k,j}^z}{2}\right) = \sigma_{m_j+1,j}^{+} \sigma_{m_j,j}^{-}\prod_{k \neq m_j,m_{j+1}}^{d-1} \sigma_{k,j}^{-} \sigma_{k,j}^{+}
\end{gather}
Restricting to $\mathcal{H}_S$, we get, 
\begin{gather}
    |m_j+1\rangle \langle m_j| = \sigma_{m_j+1,j}^{+} \sigma_{m_j,j}^{-} \Rightarrow S^+ = \sum_{m_j=0}^{d} \sigma_{m_j+1,j}^{+} \sigma_{m_j,j}^{-}
\end{gather}
where the index $m_j=d+1$ is identified with $m_j=0$ and the sum contains $d+1$ terms to account for the periodicity of the momentum states. Therefore, in the bosonic encoding, the projector $|m_j + 1\rangle \langle m_j|$ is equivalent to lowering the spin corresponding to state $m_j$ and raising the spin corresponding to state $m_j + 1$. Similarly, for the decrement operator, we get, 
\begin{equation}
    S^{-} = \sum_{m_j=0}^{d} \sigma_{m_j,j}^{+} \sigma_{m_j+1,j}^{-}
\end{equation}
Finally, we can express the interaction operators $x_ix_j$, $y_iy_j$ and $(x_iy_j + x_jy_i)$ using this encoding, 
\begin{gather}
    x_ix_j = \frac{1}{4}\sum_{m_i,m_j=0}^{d} \sigma_{m_i+1,i}^{+} \sigma_{m_i,i}^{-} \sigma_{m_j+1,j}^{+} \sigma_{m_j,j}^{-} + \sigma_{m_i+1,i}^{+} \sigma_{m_i,i}^{-} \sigma_{m_j,j}^{+} \sigma_{m_j+1,j}^{-} + h.c. \\
    y_iy_j = \frac{1}{4}\sum_{m_i,m_j=0}^{d} \sigma_{m_i+1,i}^{+} \sigma_{m_i,i}^{-} \sigma_{m_j,j}^{+} \sigma_{m_j+1,j}^{-} - \sigma_{m_i+1,i}^{+} \sigma_{m_i,i}^{-} \sigma_{m_j+1,j}^{+} \sigma_{m_j,j}^{-} + h.c. \\
    (x_iy_j + x_jy_i) = \frac{i}{2} \sum_{m_i,m_j=0}^{d} \sigma_{m_i,i}^{+} \sigma_{m_i+1,i}^{-} \sigma_{m_j,j}^{+} \sigma_{m_j+1,j}^{-} + h.c.
\end{gather}

\subsection{Sparse Linear Operator Implementation of the Bosonic Encoding Hamiltonian}

The bosonic mapping results in a large Hilbert space dimension even for small systems. Consequently, we use linear operators for our sparse diagonalization simulations which can be constructed using the action of the Hamiltonian on an arbitrary vector in the lexicographically ordered computational qubit basis. This action for the bosonic encoding is given by,  
\begin{gather}
    (H_0 v)_t = ((i-l)^2 + (j-l)^2 + (k-l)^2) v_t, \\ 
    (H_1 v)_t = \frac{3}{4} (v_{a}  + v_{b} + v_{c} + v_{d}) - \frac{1}{4} (v_{e} + v_{f} + v_{g} + v_{h})
\end{gather}
Here, $H_0$ is the diagonal angular momentum operator and $H_1$ is the pairwise nearest neighbor dipole-dipole interaction. The above equations define the non-zero part of the action. The index $t$ is constructed using the individual rotor indices $0 \leq i,j,k \leq d-1$ and is given by, 
\begin{equation}
    t = 2^{i + 2d} + 2^{j + d} + 2^{k} 
\end{equation}
For all $t' \neq t$, $(Hv)_{t'} = 0$. The indices $a$ to $d$ correspond to either raising or lowering both indices associated with a single pair of rotors, 
\begin{gather}
    a = 2^{(i-1) \% d + 2d} + 2^{(j-1)\%d + d} + 2^{k}, \\ b = 2^{(i+1) \% d + 2d} + 2^{(j+1)\%d + d} + 2^{k}, \\ c = 2^{i + 2d} + 2^{(j-1)\%d + d} + 2^{k-1}, \\ d = 2^{i + 2d} + 2^{(j+1)\%d + d} + 2^{k+1}
\end{gather}
Finally, the indices $e$ to $h$ correspond to raising one and lowering the other index associated with one pair of rotors, 
\begin{gather}
    e = 2^{(i-1) \% d + 2d} + 2^{(j+1)\%d + d} + 2^{k}, \\ f = 2^{(i+1) \% d + 2d} + 2^{(j-1)\%d + d} + 2^{k},  \\ g = 2^{i + 2d} + 2^{(j-1)\%d + d} + 2^{k+1}, \\ h = 2^{i + 2d} + 2^{(j+1)\%d + d} + 2^{k-1}
\end{gather}
The case for $N=2$ can be constructed analogously.

\end{document}